\documentclass[11pt,a4paper]{article}

\usepackage{geometry}
\usepackage{amsmath}
\usepackage{libertinus}
\usepackage{microtype}

\usepackage{booktabs}
\usepackage{array}
\usepackage{multirow}
\usepackage{graphicx}
\usepackage{xcolor}
\usepackage{enumitem}
\usepackage{pifont}
\setlist{leftmargin=1.4em,topsep=2pt,itemsep=1.5pt,parsep=0pt}

\usepackage{tikz}
\usepackage{pgfplots}

\pgfplotsset{compat=1.18}
\usepgfplotslibrary{groupplots}
\usetikzlibrary{arrows.meta,positioning,calc,decorations.pathreplacing,%
                decorations.pathmorphing,fit,backgrounds,shapes.geometric,%
                shapes.misc,shapes.multipart}

\definecolor{fbnavy}{RGB}{26,54,93}     
\definecolor{fbblue}{RGB}{32,96,160}    
\definecolor{fbred}{RGB}{178,58,38}     
\definecolor{fbgreen}{RGB}{33,113,72}   
\definecolor{fbamber}{RGB}{222,140,40}  
\definecolor{fbboxbg}{RGB}{245,247,250} 
\definecolor{fbgray}{RGB}{120,120,120}  

\tikzset{>={Stealth[length=4pt,width=3pt]}}

\pgfplotsset{
  fbbase/.style={
    axis lines=left,
    axis line style={draw=black!55, line width=0.5pt,
                     -{Stealth[length=5pt,width=3pt]}},
    tick style={black!45, line width=0.4pt},
    tick align=outside,
    major tick length=0.09cm,
    tick label style={font=\scriptsize, color=black},
    label style={font=\scriptsize, color=black},
    ymajorgrids=true,
    major grid style={draw=black!15, line width=0.35pt},
    every axis title/.append style={font=\bfseries\small, color=fbnavy},
    enlarge y limits=false,
    legend cell align=left,
    legend style={draw=none, font=\scriptsize},
  },
}

\tikzset{
  fbbox/.style={draw=black!45, rounded corners=2pt, fill=fbboxbg,
                inner sep=3pt, align=center, font=\footnotesize},
  fbnav/.style={draw=fbnavy!55, fill=fbnavy!8,  rounded corners=2pt},
  fbgrn/.style={draw=fbgreen!60, fill=fbgreen!10, rounded corners=2pt},
  fbblb/.style={draw=fbblue!60,  fill=fbblue!10,  rounded corners=2pt},
  fbredb/.style={draw=fbred!65,  fill=fbred!10,   rounded corners=2pt},
  fbar/.style={-{Stealth[length=5pt,width=3.4pt]}, draw=black!60,
               line width=0.8pt},
  fblbl/.style={font=\scriptsize, color=black!75, inner sep=1.5pt},
}

\usepackage{caption}
\usepackage[section]{placeins}

\usepackage{fancyhdr}
\usepackage{titlesec}
\titleformat{\section}{\sffamily\large\bfseries\color{fbnavy}}{\thesection}{0.6em}{}
\titleformat{\subsection}{\sffamily\normalsize\bfseries\color{fbnavy}}{\thesubsection}{0.5em}{}
\titleformat{\subsubsection}{\sffamily\small\bfseries\color{fbnavy!90!black}}{\thesubsubsection}{0.5em}{}
\titlespacing*{\section}{0pt}{1.35em}{0.45em}
\titlespacing*{\subsection}{0pt}{1.0em}{0.3em}
\titlespacing*{\subsubsection}{0pt}{0.8em}{0.25em}
\usepackage[colorlinks=true,linkcolor=fbnavy,urlcolor=fbblue,citecolor=fbblue]{hyperref}

\makeatletter
\g@addto@macro\UrlBreaks{\do\a\do\b\do\c\do\d\do\e\do\f\do\g\do\h\do\i\do\j\do\k\do\l\do\m\do\n\do\o\do\p\do\q\do\r\do\s\do\t\do\u\do\v\do\w\do\x\do\y\do\z}
\makeatother

\newcommand{\gbs}{\,GB/s}
\newcommand{\code}[1]{{\small\texttt{#1}}}
\newcommand{\fa}{FabricArena}
\newcommand{\fl}{FlashLoad}
\newcommand{\fc}{FlashClone}
\newcommand{\fb}{FlashBoot}
\title{\vspace{-1.2em}{\sffamily\bfseries\color{fbnavy}\LARGE FlashBoot: Sub-Second Weight Loading\\[2pt]
for Large Models at Rack Scale}}
\author{Isaac Zhu, Hscos Zhang, Keith Jiang, Jack Li, Hugh Yin, Jason Zhao\thanks{Corresponding author.}\\[5pt]
Scitix.AI}
\date{}

\begin{document}
\maketitle
\thispagestyle{fancy}
\vspace{-2.2em}

\begin{abstract}
\noindent
Flagship Mixture-of-Experts (MoE) models are growing along two axes at
once---total parameter count and the number of experts---with DeepSeek-V4-Pro
reaching $1.6$\,T parameters across $384$ experts. In elastic deployment scenarios
such as cold start, autoscaling, and fault recovery, many GPUs across many nodes
must become serving-ready quickly, and this growth makes weight loading a noticeable
part of the latency budget. Even on NVIDIA's GB300 NVL72, whose chip-to-chip (C2C) link and NVLink Switch
fabric have ample raw bandwidth, today's state-of-the-art loaders (Hugging~Face
SafeTensors, InstantTensor, and SGLang's NCCL-based Tensor R-Fork) leave most of that
bandwidth unused. The losses are structural: \textbf{(C1)} weight memory is
fragmented into tens of thousands of per-tensor objects, so transfers run far below
link bandwidth; \textbf{(C2)} cross-node replication is gated by NCCL communicator
setup, which costs $10$--$110$\,s before a single weight byte moves; and
\textbf{(C3)} the existing cross-node GPU$\to$GPU clone path is serial and scales
poorly to concurrent multi-node bring-up.

\textbf{We present \fb, a hardware-friendly, framework--workflow co-designed weight-loading subsystem built on SGLang.}
At its core is \fa, a contiguous, exportable device-memory layout that lets a model's
discrete weight tensors occupy one large 1-D image while remaining inter-node
addressable. On top of it, \fl{} loads from CPU (host$\to$GPU) as a single bulk,
zero-copy transfer, and \fc{} replicates a resident model GPU$\to$GPU by a
remote-mapping mechanism that removes NCCL setup.
In experiments on NVL72 with DeepSeek-V4-Pro ($1.6$\,T) and DeepSeek-V4-Flash
($284$\,B), \fc{} maps remote weight memory in $\sim$$10$\,ms (versus $10$--$110$\,s
for NCCL) and sustains $\geq$$700$\gbs{} per clone. Against the state of the art,
\fb{} accelerates single-node weight loading by up to $\mathbf{50\times}$ (from
$20.1$\,s to $0.4$\,s) and concurrent rack-level weight loading by
$\mathbf{>270\times}$ (from $87$\,s to $0.32$\,s).
\end{abstract}

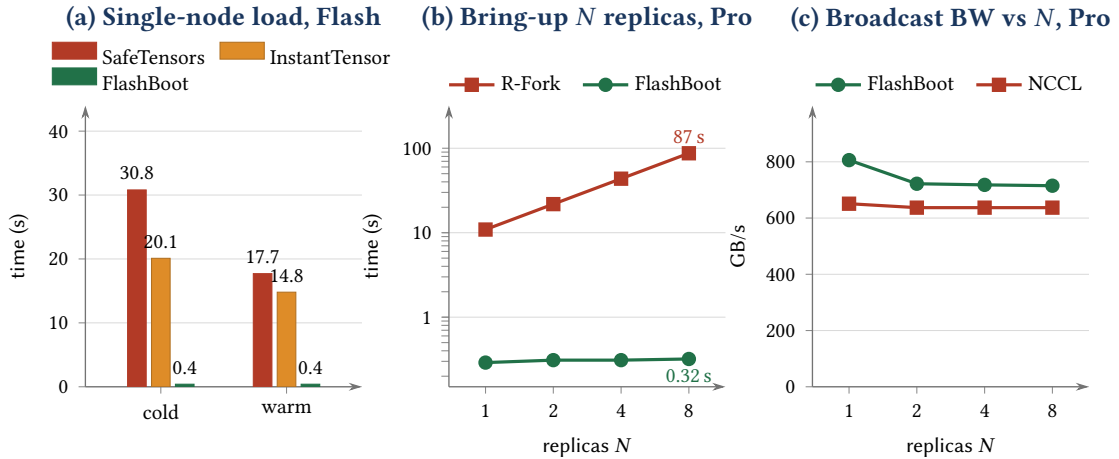
\begin{figure}[!ht]
\centering
\begin{tikzpicture}
\begin{groupplot}[
  group style={group size=3 by 1, horizontal sep=1.15cm},
  width=0.31\textwidth, height=5.3cm, fbbase, ymin=0,
  title style={yshift=18pt, font=\bfseries\small\color{fbnavy}},
  every tick label/.append style={font=\scriptsize,color=black},
  label style={font=\scriptsize\sffamily,color=black},
]
\nextgroupplot[ybar, bar width=7pt, title={(a) Single-node load, Flash},
  ylabel={time (s)}, symbolic x coords={cold,warm}, xtick=data,
  xticklabels={cold,warm}, ymax=44, enlarge x limits=0.6,
  nodes near coords, nodes near coords style={font=\scriptsize,/pgf/number format/fixed,/pgf/number format/precision=1},
  legend style={at={(0.5,1.0)},anchor=south,legend columns=2,draw=none,font=\scriptsize,row sep=-2.5pt,
    /tikz/every even column/.append style={column sep=0.12cm}},
  legend image code/.code={\draw[#1](0,0.0)rectangle(0.16,0.06);}]
\addplot[fill=fbred,draw=fbred] coordinates {(cold,30.8)(warm,17.7)}; \addlegendentry{SafeTensors}
\addplot[fill=fbamber,draw=fbamber!80!black] coordinates {(cold,20.1)(warm,14.8)}; \addlegendentry{InstantTensor}
\addplot[fill=fbgreen,draw=fbgreen] coordinates {(cold,0.4)(warm,0.4)}; \addlegendentry{FlashBoot}
\nextgroupplot[title={(b) Bring-up $N$ replicas, Pro}, ymode=log,
  ylabel={time (s)}, symbolic x coords={1,2,4,8}, xtick=data, enlarge x limits=0.18,
  xlabel={replicas $N$}, ymin=0.15, ymax=320, ytick={0.1,1,10,100},
  yticklabels={0.1,1,10,100},
  every axis plot/.append style={very thick,mark size=2.1pt},
  legend style={at={(0.5,1.0)},anchor=south,legend columns=-1,draw=none,font=\scriptsize,
    /tikz/every even column/.append style={column sep=0.2cm}}]
\addplot[fbred,mark=square*] coordinates {(1,10.9)(2,21.8)(4,43.6)(8,87.2)}; \addlegendentry{R-Fork}
\addplot[fbgreen,mark=*] coordinates {(1,0.29)(2,0.31)(4,0.31)(8,0.32)}; \addlegendentry{FlashBoot}
\node[font=\scriptsize,color=fbred,anchor=south] at (axis cs:8,87.2) {87\,s};
\node[font=\scriptsize,color=fbgreen,anchor=north] at (axis cs:8,0.32) {0.32\,s};
\nextgroupplot[title={(c) Broadcast BW vs $N$, Pro}, ylabel={GB/s},
  symbolic x coords={1,2,4,8}, xtick=data, enlarge x limits=0.18,
  xlabel={replicas $N$}, ymin=0, ymax=1000, ytick={0,200,400,600,800},
  every axis plot/.append style={very thick,mark size=2.1pt},
  legend style={at={(0.5,1.0)},anchor=south,legend columns=-1,draw=none,font=\scriptsize,
    /tikz/every even column/.append style={column sep=0.2cm}}]
\addplot[fbgreen,mark=*] coordinates {(1,806)(2,722)(4,718)(8,715)}; \addlegendentry{FlashBoot}
\addplot[fbred,mark=square*] coordinates {(1,651)(2,637)(4,637)(8,637)}; \addlegendentry{NCCL}
\end{groupplot}
\end{tikzpicture}
\caption{\textbf{Results at a glance.} \textbf{(a)}~Single-node DeepSeek-V4-Flash load time:
FlashBoot collapses the engine-visible load to $\sim$$0.4$\,s versus
$15$--$31$\,s for SafeTensors and InstantTensor. \textbf{(b)}~Time to bring up
$N$ DeepSeek-V4-Pro replicas: FlashBoot fills all clones concurrently along a chain in
$\sim$$0.32$\,s, whereas SGLang's R-Fork stands up an NCCL group and copies
serially, so its wall-clock grows with $N$ (it serializes the per-clone standup and transfer; \S\ref{sec:e-clone}).
\textbf{(c)}~Chain-broadcast bandwidth per clone stays near the link rate as $N$
grows and exceeds NCCL broadcast on the same fabric.}
\label{fig:summary}
\end{figure}

\clearpage
\section{Introduction and Background}\label{sec:intro}

\subsection{Where fast weight loading matters}
Outside steady-state serving, an inference service loads its weights more than
once, and sometimes under a deadline, in several elastic-deployment scenarios:

\begin{itemize}
  \item \textbf{Initial deployment / cold start.} A fresh service brings a model
  onto a pool of GPUs from storage; serving cannot begin until the last byte is
  resident.
  \item \textbf{Serverless / scale-to-zero.} A model that is paged out to reclaim
  capacity must be paged back in on the next request, so load latency lands
  directly on the request's tail.
  \item \textbf{Autoscaling.} When load rises, the scheduler adds replicas; the
  faster a new replica becomes serving-ready, the tighter the system can track
  demand instead of over-provisioning for it.
  \item \textbf{Fault recovery.} When a node or a process dies, the service must
  restart it or stand up a replacement and re-fill its weights to restore
  capacity and redundancy.
\end{itemize}

\noindent These scenarios share one path: getting a (often very large) weight
image onto one or many GPUs, frequently across several nodes at once, quickly. As
models grow this path lengthens, and in latency-sensitive cases it can slow how
quickly a deployment reacts. FlashBoot targets this path in these scenarios; it
does not change steady-state serving.

\subsection{Two axes of model growth}
\begin{figure}[t]
\centering
\begin{tikzpicture}
\begin{groupplot}[
  group style={group size=2 by 1, horizontal sep=1.7cm},
  width=0.49\textwidth, height=4.7cm, fbbase, ymin=0,
  xtick=data, enlarge x limits=0.08,
  x tick label style={font=\tiny,rotate=38,anchor=north east,color=black},
  nodes near coords, nodes near coords style={font=\tiny,/pgf/number format/fixed,/pgf/number format/precision=0},
  every axis plot/.append style={draw=none},
]
\nextgroupplot[ybar, bar width=8pt, title={(a) total parameters (B)}, ylabel={parameters (B)},
  symbolic x coords={GPT-oss,Flash,Qwen3.5,Llama4,Nemotron3,GLM-5.1,Kimi-2.6,Pro,Kimi-K3},
  ymax=3200, ytick={0,800,1600,2400}]
\addplot[fill=fbblue] coordinates {(GPT-oss,120)(Flash,284)(Qwen3.5,397)(Llama4,400)(Nemotron3,550)(GLM-5.1,754)(Kimi-2.6,1000)(Pro,1600)(Kimi-K3,2800)};
\nextgroupplot[ybar, bar width=8pt, title={(b) experts per MoE layer}, ylabel={experts},
  symbolic x coords={GPT-oss,Llama4,Qwen3.5,GLM-5.1,Flash,Kimi-2.6,Pro,Nemotron3,Kimi-K3},
  ymax=1080, ytick={0,256,512,768}]
\addplot[fill=fbnavy!72] coordinates {(GPT-oss,128)(Llama4,128)(Qwen3.5,160)(GLM-5.1,160)(Flash,256)(Kimi-2.6,384)(Pro,384)(Nemotron3,512)(Kimi-K3,896)};
\end{groupplot}
\end{tikzpicture}
\caption{\textbf{Two axes of growth in flagship open-weight MoE models.}
\textbf{(a)}~total parameters (ordered low to high) and \textbf{(b)}~experts per
MoE layer (ordered low to high; Kimi-K3 has the most at $896$) both trend
upward. Values are approximate, from public model cards and
reports~\cite{gptoss,llama4,qwen,glm,nemotron,kimi,kimik3,deepseek}; the two
DeepSeek-V4 models used in this report (Flash, $284$\,B / $256$ experts; Pro,
$1.6$\,T / $384$) sit on this trend, and the recently announced Kimi-K3
($2.8$\,T / $896$ experts) now tops both axes. More experts is what turns a checkpoint into
tens of thousands of small per-tensor objects; more parameters is what makes the
aggregate byte volume large.}
\label{fig:trend}
\end{figure}
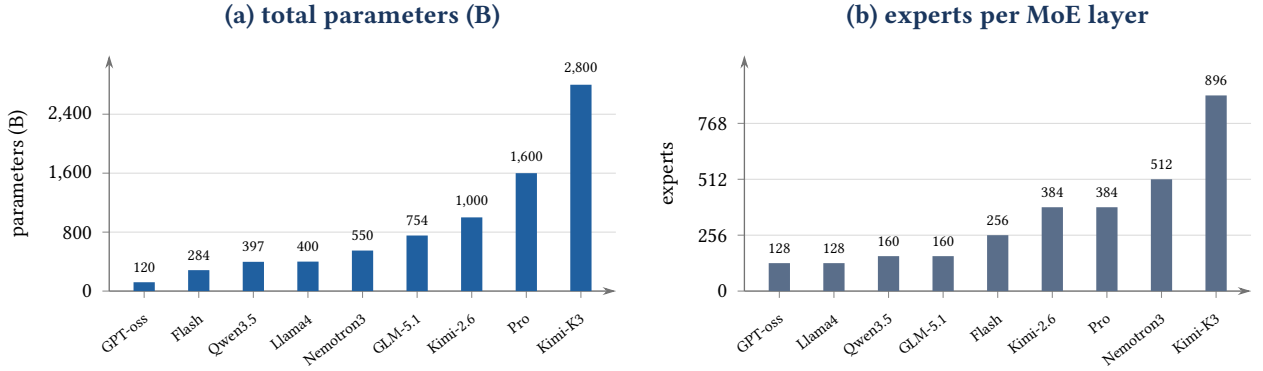

Open-weight flagship models are growing along two axes simultaneously
(Figure~\ref{fig:trend}). The first is total parameter count, which has climbed
from the hundred-billion class to beyond a trillion in roughly a year. The
second, specific to the Mixture-of-Experts (MoE) designs that now dominate the
frontier, is the number of experts: each layer fans out into many expert
weight tensors, of which only a few are active per token. The forthcoming
DeepSeek-V4-Pro reaches $1.6$\,T parameters across $384$ experts. The two axes
compound: the expert dimension is precisely what multiplies a checkpoint into
tens of thousands of individual weight tensors, and the parameter axis is
what makes the aggregate byte volume large. Both directly inflate weight
loading: one by sheer volume, the other by sheer object count.

\subsection{The thesis}
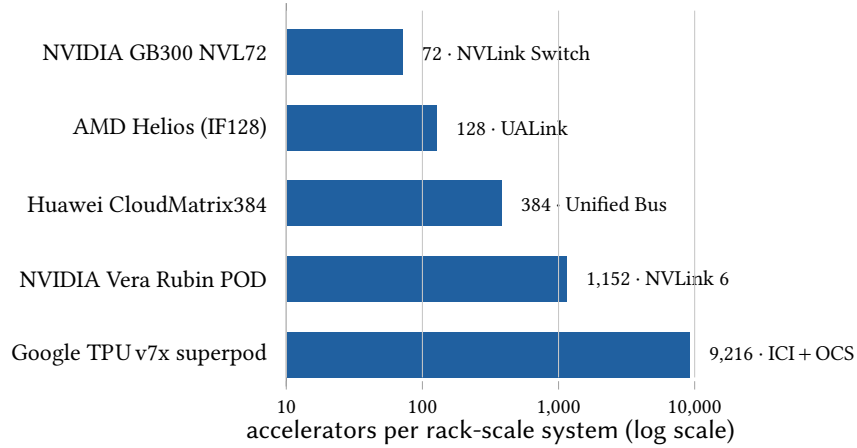
\begin{figure}[t]
\centering
\begin{tikzpicture}[font=\small]
\def\k{1.8}\def\xmn{1.0}
\foreach \y/\v/\vl/\lab in {%
  4/72/{72\ $\cdot$\ NVLink Switch}/NVIDIA GB300 NVL72,%
  3/128/{128\ $\cdot$\ UALink}/AMD Helios (IF128),%
  2/384/{384\ $\cdot$\ Unified Bus}/Huawei CloudMatrix384,%
  1/1152/{1{,}152\ $\cdot$\ NVLink 6}/NVIDIA Vera Rubin POD,%
  0/9216/{9{,}216\ $\cdot$\ ICI\,$+$\,OCS}/Google TPU\,v7x superpod}{
  \pgfmathsetmacro\w{(log10(\v)-\xmn)*\k}
  \fill[fbblue] (0,\y) rectangle (\w,\y+0.6);
  \node[anchor=east,font=\footnotesize,color=black] at (-0.12,\y+0.3) {\lab};
  \node[anchor=west,font=\scriptsize,color=black] at (\w+0.12,\y+0.3) {\vl};
}
\draw[black!40,line width=0.5pt] (0,-0.12)--(0,4.95);
\foreach \d/\t in {1/10,2/100,3/{1{,}000},4/{10{,}000}}{
  \pgfmathsetmacro\xx{(\d-\xmn)*\k}
  \draw[black!22,line width=0.4pt] (\xx,0)--(\xx,4.85);
  \draw[black!35] (\xx,-0.05)--(\xx,-0.16) node[below,font=\scriptsize,color=black]{\t};
}
\node[font=\small\sffamily,color=black] at (2.7,-0.72) {accelerators per rack-scale system (log scale)};
\end{tikzpicture}
\caption{\textbf{Rack-scale systems are the emerging unit of large-model
deployment.} Tightly-interconnected accelerators per rack-scale system across
vendors, ordered smallest to largest (log scale): NVIDIA GB300 NVL72 ($72$
GPUs)~\cite{gb300}, AMD Helios IF128 ($128$ GPUs)~\cite{helios}, Huawei
CloudMatrix384 ($384$ Ascend NPUs)~\cite{cloudmatrix}, NVIDIA Vera Rubin POD
($1{,}152$ Rubin GPUs)~\cite{verarubin}, and a Google TPU\,v7x (Ironwood) superpod
(up to $9{,}216$ chips)~\cite{tpu}. The reported unit differs, single
NVLink/UALink/Unified-Bus scale-up domains versus multi-rack pods, but the trend is
uniform: the deployment unit has grown from tens to thousands of
high-bandwidth-coupled accelerators. We evaluate \fb{} on the GB300 NVL72, the
smallest and most widely available member of the class.}
\label{fig:rackscale}
\end{figure}

The hardware industry has answered the bandwidth side of this problem with a
structural shift: the \emph{rack-scale system} is becoming the unit of large-model
deployment. A rack is no longer a loose set of servers behind a commodity network
but a co-designed machine in which tens of accelerators share a high-bandwidth,
near-all-to-all ``scale-up'' fabric, with a separate scale-out network between
racks. NVIDIA's GB300 NVL72 binds $72$ Blackwell-Ultra GPUs over an NVLink Switch
fabric~\cite{gb300}, and its successor, the Vera Rubin POD, carries the same idea
across five co-designed rack types~\cite{verarubin}; the pattern is industry-wide,
spanning AMD's Helios rack (UALink-over-Ethernet scale-up)~\cite{helios}, Huawei's
CloudMatrix384 supernode ($384$ Ascend NPUs on an all-to-all Unified
Bus)~\cite{cloudmatrix}, and Google's TPU pods (thousands of chips on a 3-D-torus
ICI fabric)~\cite{tpu}; across them the deployment unit spans from tens to thousands
of high-bandwidth-coupled accelerators (Figure~\ref{fig:rackscale}). Two forces
drive this convergence: the two axes of model
growth above, and the \emph{token economics} of inference, the explosion in tokens
consumed by reasoning and agentic workloads, which together reward packing more
model into one high-bandwidth domain and standing serving replicas up fast enough
to track demand. Fast, elastic weight loading is precisely the operation that turns
this dense fabric into served capacity on a deadline.

On a rack-scale GB300 NVL72, every GPU reaches host memory over a fast
Grace$\leftrightarrow$GPU C2C link and reaches every other GPU in the rack
over an all-to-all NVLink Switch fabric (\S\ref{sec:hw}). On paper, that is more
than enough bandwidth to fill a model in well under a second. Yet, as we show in
\S\ref{sec:challenges} with direct measurements, today's state-of-the-art loaders
realize only a small fraction of it: a Pro rank that the C2C bus could move in
$\sim$$1.2$\,s still takes $\sim$$45$--$117$\,s to load, and replicating a resident
model to even one other node first pays a $10$\,s$+$ NCCL tax before any weight
byte moves. The gap is not bandwidth; it is how the bytes are laid out and
moved. This report makes that case quantitatively and then presents \fb{}, a
hardware-friendly framework--workflow co-design that closes the gap by changing
the weight memory layout first and rebuilding the load and replication
paths on top of it. We study the GB300 NVL72 as the most widely available member
of this class, but the design rests only on properties these systems share, a
contiguous, exportable memory region and a fast intra-rack fabric, so its ideas
carry to the others (\S\ref{sec:beyond}).

\section{Opportunity and Challenges}\label{sec:challenges}

\subsection{The rack-scale hardware opportunity}\label{sec:hw}
The rack-scale systems surveyed in \S\ref{sec:intro} share one structural property
that fast loading can exploit: a high-bandwidth, near-all-to-all intra-rack
scale-up fabric that is far faster than each accelerator's path to host memory.
Unless stated otherwise, experiments in this report run on a GB300 NVL72 system
(NVIDIA Grace--Blackwell Ultra), which we take as the most widely available exemplar
of this class; a study of other rack-scale platforms (e.g.\ H100 islands joined by
RDMA) is left to future work and sketched in \S\ref{sec:beyond}.
Table~\ref{tab:env} summarizes the platform and
Table~\ref{tab:bw} its interconnect bandwidths, giving both the vendor specification and the
values we measure on the machine. Two links matter for loading: the
Grace$\leftrightarrow$GPU \textbf{C2C} link (the host$\to$GPU path) and the
GPU$\leftrightarrow$GPU \textbf{NVLink} fabric (the GPU-to-GPU path). The
structural fact we lean on throughout is that per-GPU NVLink bandwidth is
$\sim$$4\times$ the per-GPU C2C bandwidth; this is the hardware reason that, once
a model is resident somewhere in the rack, replicating it GPU$\to$GPU is
fundamentally cheaper than reloading it from host memory.

\begin{table}[t]
\centering\small
\caption{Platform and software environment.}
\label{tab:env}
\begin{tabular}{@{}l >{\raggedright\arraybackslash}p{9.4cm}@{}}
\toprule
Item & Value \\
\midrule
System & GB300 NVL72 (NVIDIA Grace--Blackwell Ultra) \\
GPUs per node & 4 (\code{sm\_103}); 2 Grace CPUs per node \\
Rack & one NVLink-Switch fabric domain (single IMEX partition) \\
Nodes used & up to 9 (1 seed $+$ 8 clones) of a $\sim$17-node rack \\
NVIDIA driver & \code{580.105.08} \\
CUDA / runtime & \code{13.0} (nvcc 13.0.88), \code{torch 2.9.1+cu130}, NCCL \code{2.27.7} \\
\bottomrule
\end{tabular}
\end{table}

\begin{table}[t]
\centering\small
\caption{GB300 NVL72 interconnect bandwidths (vendor spec vs.\ measured).}
\label{tab:bw}
\begin{tabular}{@{}l r r@{}}
\toprule
Path & Spec (per dir.) & Measured \\
\midrule
NVLink, per GPU            & 1.8\,TB/s ($\approx$900) & 825--826\gbs{} (1 copy) \\
\quad seed-GPU egress      & ---                      & 837--839\gbs{} \\
NVLink Switch fabric       & 130\,TB/s                & --- \\
Grace--GPU C2C, per GPU    & 450\,GB/s ($\approx$225) & 223.5\gbs{} \\
\quad H2D (effective)      & ---                      & $\sim$185\gbs{} \\
HBM3e, per GPU             & 8\,TB/s                  & --- \\
LPDDR5X, per Grace         & ---                      & $\sim$500\gbs{} \\
\bottomrule
\end{tabular}
\end{table}

\noindent\textbf{Models under test.} We use the two checkpoints in
Table~\ref{tab:models}: Flash (DeepSeek-V4-Flash) and the much larger
Pro (DeepSeek-V4-Pro). Both are evaluated at tensor-parallel degree
$\text{TP}{=}4$ (one shard per GPU on a node); we use TP as the running example
but the design supports the parallelism modes inference engines use in general
(\S\ref{sec:design}). Per-rank sizes are byte-exact: a Pro rank is
$217.0$\,GB, namely $205.51$\,GB of expert weights plus $11.44$\,GB of non-expert
weights. The per-rank total times four exceeds the on-disk checkpoint because
non-expert weights are replicated across TP ranks and the pre-packed image carries
the kernel-ready, post-quantization layout rather than the compact on-disk form.

\begin{table}[t]
\centering\small
\caption{Evaluation models (DeepSeek-V4 family) at TP4.}
\label{tab:models}
\begin{tabular}{@{}l r r r r r@{}}
\toprule
Model & Params & Experts & Layers & Checkpoint & Per-rank (expert\,+\,non-expert) \\
\midrule
Flash & 284\,B & 256 & 43 & 149\,GB & 41.4\,GB \;($36.8+4.64$) \\
Pro   & 1.6\,T & 384 & 61 & 805\,GB & 217.0\,GB \;($205.51+11.44$) \\
\bottomrule
\end{tabular}
\end{table}

\subsection{The baseline reality}\label{sec:baseline}
\begin{figure}[t]
\centering
\begin{tikzpicture}
\begin{groupplot}[
  group style={group size=2 by 1, horizontal sep=1.7cm},
  width=0.49\textwidth, height=4.7cm, fbbase, ymin=0,
  symbolic x coords={cold,warm}, xtick=data, enlarge x limits=0.55,
  nodes near coords,
  nodes near coords style={font=\scriptsize,/pgf/number format/fixed,/pgf/number format/precision=1},
  xticklabels={cold (disk),warm (RAM)}, title style={yshift=12pt},
]
\nextgroupplot[ybar, bar width=12pt, title={(a) Flash 149\,GB, TP4}, ylabel={engine load time (s)},
  ymax=40,
  legend columns=-1,
  legend style={at={(0.5,1.0)},anchor=south,draw=none,font=\scriptsize,
    /tikz/every even column/.append style={column sep=0.35cm}},
  legend image code/.code={\draw[#1](0,0.0)rectangle(0.05,0.07);}]
\addplot[fill=fbred,draw=fbred] coordinates {(cold,30.8)(warm,17.7)}; \addlegendentry{SafeTensors}
\addplot[fill=fbamber,draw=fbamber!80!black] coordinates {(cold,20.1)(warm,14.8)}; \addlegendentry{InstantTensor}
\nextgroupplot[ybar, bar width=12pt, title={(b) Pro 805\,GB, TP4}, ylabel={engine load time (s)},
  ymax=135]
\addplot[fill=fbred,draw=fbred] coordinates {(cold,116.8)(warm,67.4)};
\addplot[fill=fbamber,draw=fbamber!80!black] coordinates {(cold,72.9)(warm,45.8)};
\end{groupplot}
\end{tikzpicture}
\caption{\textbf{State-of-the-art single-node load time} (engine-visible weight
load, TP4). Even from warm RAM (disk removed), a Pro rank takes $45$--$67$\,s,
yet its $217$\,GB would cross the C2C bus in $\sim$$1.2$\,s at $185$\gbs. The
loaders are bus-idle $>97\%$ of the time; the cost is per-tensor overhead, not
bandwidth (C1).}
\label{fig:baseline}
\end{figure}
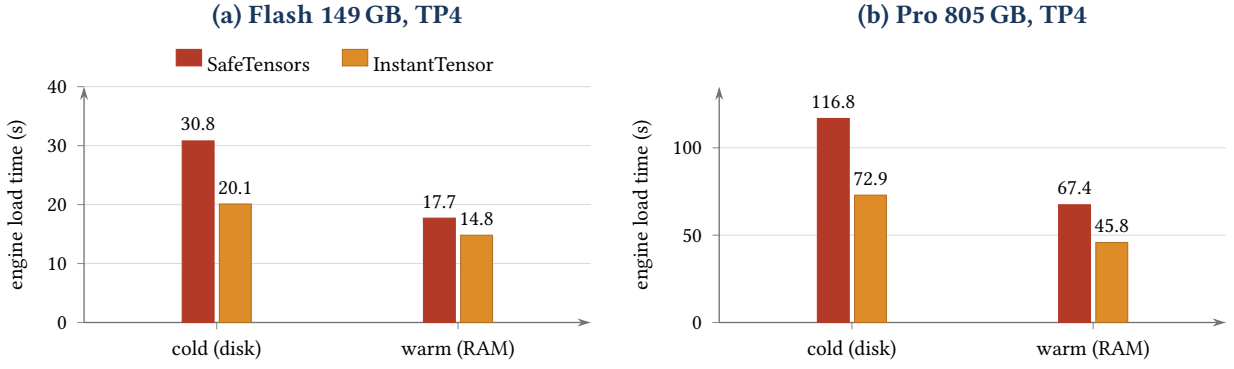

We measured three representative loaders on this hardware: Hugging~Face
SafeTensors (the de-facto standard loader), the academic
InstantTensor, and SGLang's production Tensor R-Fork
(\code{load\_from\_remote\_instance}, an NCCL-based GPU$\to$GPU replication path).
Figure~\ref{fig:baseline} shows single-node load time for SafeTensors and
InstantTensor from cold disk and from warm RAM (\code{/dev/shm}). Even from warm
RAM (with disk I/O entirely removed), a Pro rank still takes
$45$--$67$\,s and a Flash rank $\sim$$15$--$18$\,s. Since the C2C bus can move a Pro
rank's $217$\,GB in $\sim$$1.2$\,s at the measured $\sim$$185$\gbs, the loaders
are spending $\mathbf{>97\%}$ of their time not moving bytes over the bus.
Something other than bandwidth dominates. We now isolate the three structural
causes; each is stated as a claim, derived, and backed with a measured number,
and each maps to a specific element of \fb's design (\S\ref{sec:design}).

\subsection{C1: fragmented weight memory starves the interconnect}\label{sec:c1}
The standard loader treats a checkpoint as what it is on disk: a large
collection of independent tensors. Each parameter, and for an MoE model,
each expert, is allocated and transferred as its own object, with re-layout and
staging copies along the way. A Pro rank has on the order of tens of thousands of
such objects (e.g.\ $384$ experts across $61$ layers). The dominant cost is then
the count of operations, not
the byte volume, which is exactly why the load stays slow even when the bytes
already sit in RAM (Figure~\ref{fig:baseline}).

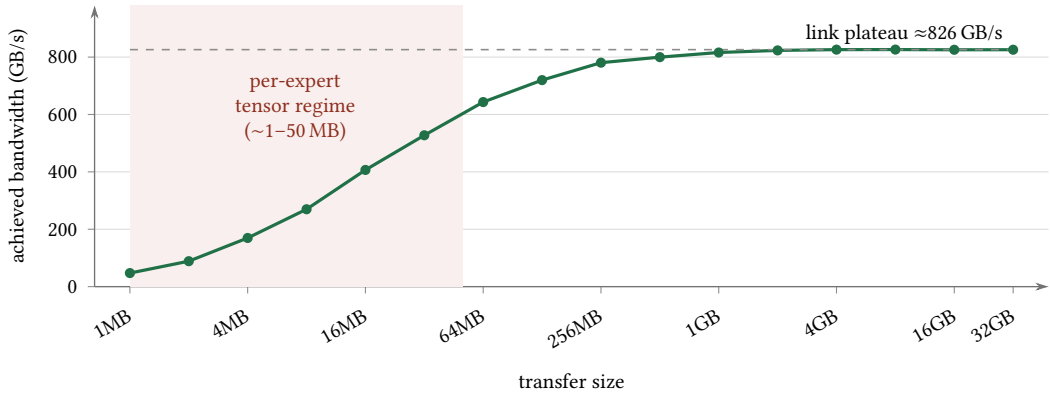
\begin{figure}[t]
\centering
\begin{tikzpicture}
\begin{axis}[fbbase, width=0.84\textwidth, height=5.3cm,
  xmode=log, log basis x=2,
  xlabel={transfer size}, ylabel={achieved bandwidth (GB/s)},
  xtick={1,4,16,64,256,1024,4096,16384,32768},
  xticklabels={1MB,4MB,16MB,64MB,256MB,1GB,4GB,16GB,32GB},
  x tick label style={font=\scriptsize,rotate=30,anchor=north east},
  ymin=0, ymax=980, ytick={0,200,400,600,800},
  enlarge x limits=0.04,
  every axis plot/.append style={very thick}]
\fill[fbred!8] (axis cs:1,0) rectangle (axis cs:50,980);
\node[font=\scriptsize,color=fbred!80!black,align=center,anchor=south]
  at (axis cs:7,470) {per-expert\\tensor regime\\($\sim$1--50\,MB)};
\addplot[fbgreen,mark=*,mark size=1.3pt] coordinates {
 (1,47.3)(2,88.7)(4,169.6)(8,269.7)(16,406.7)(32,527.5)(64,643.3)(128,719.8)
 (256,780.5)(512,799.8)(1024,816.0)(2048,823.3)(4096,826.0)(8192,826.2)
 (16384,825.5)(32768,825.7)};
\addplot[black!45,dashed,line width=0.6pt] coordinates {(1,826)(32768,826)};
\node[font=\scriptsize,color=black,anchor=north east] at (axis cs:32768,950) {link plateau $\approx$826\gbs};
\end{axis}
\end{tikzpicture}
\caption{\textbf{Small transfers starve the interconnect (C1).} Achieved
bandwidth of a single cross-node DMA versus payload size (GB300 NVL72). Peak
($\sim$$826$\gbs) is reached only past $\sim$$2$\,GB; in the per-expert-tensor
regime ($\sim$$1$--$50$\,MB) even an ideal transport sustains a fraction of the
link. Loading or replicating a model tensor-by-tensor therefore cannot use the
fabric, regardless of how fast the fabric is.}
\label{fig:frag}
\end{figure}

The bandwidth penalty of small transfers is intrinsic, not an artifact of the
loader. A single cross-node copy-engine DMA only approaches link bandwidth when
it is large: Figure~\ref{fig:frag} sweeps one transfer from 1\,MB to
32\,GB and shows the achieved bandwidth climbing from $47$\gbs{} at 1\,MB to a
$\sim$$826$\gbs{} plateau only past $\sim$$2$\,GB. Per-expert tensors live in the
$\sim$$1$--$50$\,MB regime (shaded), where even a perfect transport reaches only
$\sim$$50$--$500$\gbs, a fraction of the link. Fragmentation thus caps every
movement of the weights, on both the host$\to$GPU and the GPU$\to$GPU paths. The
implication for design is direct: make the weights contiguous so they can
move in one big transfer (\S\ref{sec:arena}).

\subsection{C2: NCCL communicator setup dominates cross-node replication}\label{sec:c2}
\begin{figure}[t]
\centering
\begin{tikzpicture}[font=\small]
\def\k{1.55}\def\xmn{0.6}
\pgfmathsetmacro\xax{(5.78-\xmn)*\k}
\foreach \y/\v/\c/\lab/\vl in {%
  3/11900/fbred/NCCL init (20 ranks)/11.9\,s,%
  2/354/fbred/NCCL group (1-to-N)/354\,ms,%
  1/112/fbred/NCCL group (pairwise)/112\,ms,%
  0/10/fbgreen/FlashBoot map/10\,ms}{
  \pgfmathsetmacro\w{(log10(\v)-\xmn)*\k}
  \fill[\c] (0,\y) rectangle (\w,\y+0.6);
  \node[anchor=east,font=\footnotesize,color=black] at (-0.12,\y+0.3) {\lab};
  \node[anchor=west,font=\scriptsize,color=black] at (\w+0.1,\y+0.3) {\vl};
}
\draw[black!40,line width=0.5pt] (0,-0.12)--(0,3.95);
\foreach \d/\t in {1/10,2/100,3/{1{,}000},4/{10{,}000},5/{100{,}000}}{
  \pgfmathsetmacro\xx{(\d-\xmn)*\k}
  \draw[black!22,line width=0.4pt] (\xx,0)--(\xx,3.85);
  \draw[black!35] (\xx,-0.05)--(\xx,-0.16) node[below,font=\scriptsize,color=black]{\t};
}
\node[font=\small\sffamily,color=black] at (\xax/2,-0.72) {one-time setup before first weight byte (ms, log scale)};
\end{tikzpicture}
\caption{\textbf{Cross-node setup cost before the first weight byte (C2),} log
scale. NCCL must stand up a communicator; the components shown reach $11.9$\,s for
\code{init\_process\_group} over 20 ranks alone, and the full standup is $10$--$110$\,s in practice. \fb{} instead imports a
$64$-byte fabric handle and maps remote memory in $\sim$$10$\,ms, a
$\sim$$10^3$--$10^4\times$ reduction, then moves the $\sim$$0.30$\,s of bytes.}
\label{fig:nccl}
\end{figure}
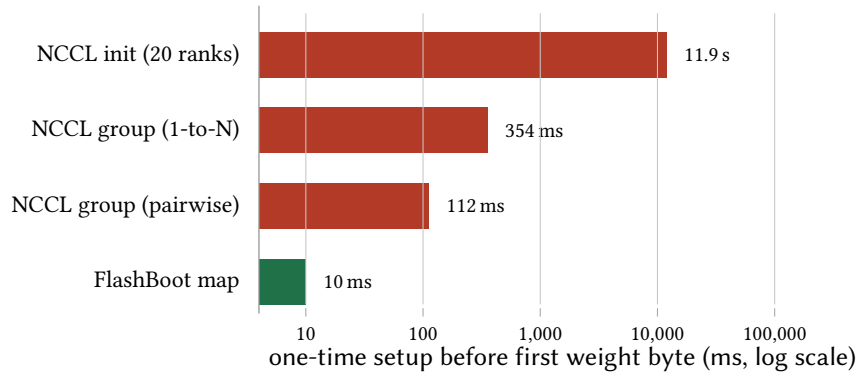

Replicating an already-resident model to another node is the natural way to
exploit the rack: read once from storage, then copy GPU$\to$GPU over the fast
fabric. SGLang's production path (R-Fork) does exactly this, but routes the copy
through a \textbf{NCCL} process group, which must be stood up before any
weight byte moves: a rendezvous, \code{ncclCommInit} (P2P buffer allocation,
per-pair connection setup, ring/tree topology search), a barrier, and an SM-kernel
warm-up. We measure this standup at $\mathbf{10}$--$\mathbf{110}$\,\textbf{s}
(e.g.\ \code{init\_process\_group} over $20$ ranks alone takes $11.9$\,s;
Figure~\ref{fig:nccl}).

To see how badly this inverts the cost, do the arithmetic on the transfer it
gates. A Pro rank is $217$\,GB; at $80\%$ of the $\sim$$900$\gbs{} per-GPU NVLink
direction, the bytes themselves cross in
\[
t_{\text{xfer}}=\frac{217\ \text{GB}}{0.8\times 900\ \text{GB/s}}\approx 0.30\ \text{s}.
\]
Setup is thus $30$--$370\times$ the transfer it enables. For a library
used continuously this amortizes; for weight loading, a one-shot or low-frequency
event, it is pure overhead on the critical path. A general-purpose collective
library is simply the wrong tool for fast startup. The design implication: replace
the communicator with a handle exchange that maps remote memory directly, in
milliseconds, with no group to build (\S\ref{sec:remotemap}).

\subsection{C3: the production path is serial and does not scale}\label{sec:c3}
The third problem appears precisely in the scenario that matters most for
elasticity: bringing up many replicas at once. R-Fork is, by construction,
a serial seed$\to$clone design, with a fresh communicator standup and transfer per
clone, so the wall-clock to fill $N$ replicas grows roughly linearly in $N$. Even
setting NCCL aside, the obvious ``parallel'' alternative of having all $N$ clones
pull from one seed at once (a 1$\to$N star) does not help, because a
single seed GPU's NVLink egress is a fixed ceiling ($837$--$839$\gbs,
measured) that the $N$ readers must share. Per-clone bandwidth then collapses as
$\sim$egress$/N$ for $N\ge2$: we measure $825$\gbs{} at $N{=}1$ (a single copy,
limited by the link rather than egress) but only $\sim$$104$\gbs{} at $N{=}8$
($\approx 838/8$; \S\ref{sec:e-clone}, Figure~\ref{fig:clone}). Either way, whether serial standup or shared
egress, the time to make a rack of replicas serving-ready scales the wrong way
with the rack size. The design implication is to remove the single-seed
bottleneck entirely with a chained, pipelined broadcast in which every node
reads from a distinct predecessor over an independent link (\S\ref{sec:topo}).

\medskip
\noindent Taken together, C1--C3 say the problem is not the GB300's bandwidth but
the layout and movement discipline layered on top of it. \fb{} addresses
all three with one coherent design, to which we now turn.

\section{Design}\label{sec:design}

\fb{} is a weight-loading subsystem for NVLink/RDMA-interconnected GPU racks. It
optimizes both halves of the loading lifecycle: loading from CPU
(\fl, host$\to$GPU) and replicating from a resident peer (\fc, GPU$\to$GPU). It is designed for rack-scale platforms such as GB300 NVL72. While our prototype
targets the NVLink Switch fabric, the design's central ideas apply equally to
NVLink-only systems and, in principle, to other accelerators with an
exportable-memory primitive. Throughout we use tensor parallelism (TP) as the
running example for concreteness, but nothing in the design is TP-specific: the
weight image is treated as an opaque byte range, so the same machinery applies to
the data-, expert-, and pipeline-parallel layouts an inference engine produces
(the coverage our prototype realizes today is detailed in \S\ref{sec:impl}).

\subsection{Design principles}
The challenges of \S\ref{sec:challenges} translate into five principles that the
rest of this section instantiates:
\begin{enumerate}[label=\textbf{P\arabic*.}]
  \item \textbf{Fix the layout first.} Contiguity is the enabler for everything
  else: a single large image can be moved in one bandwidth-saturating transfer,
  whereas tens of thousands of per-tensor objects cannot (answers C1).
  \item \textbf{One image, served in place.} The weights live exactly once and the
  model's parameters are views onto that image, with no per-parameter storage
  and no redundant same-device copies ($1\times$ memory).
  \item \textbf{Map memory, don't build a communicator.} Cross-node access is a
  one-time exchange of a small handle plus a direct memory mapping, not a
  collective-library group (answers C2).
  \item \textbf{No single-point bottleneck.} Every transfer is arranged so that
  each GPU link carries exactly one reader, so aggregate bandwidth is independent
  of how many nodes participate (answers C3).
  \item \textbf{Pay slow work once, and off the critical path.} Sharding,
  quantization, and re-layout are done once, offline; the slow cold disk read is
  overlapped with engine startup so the boot critical path is pure data movement.
\end{enumerate}

\subsection{Architecture overview}
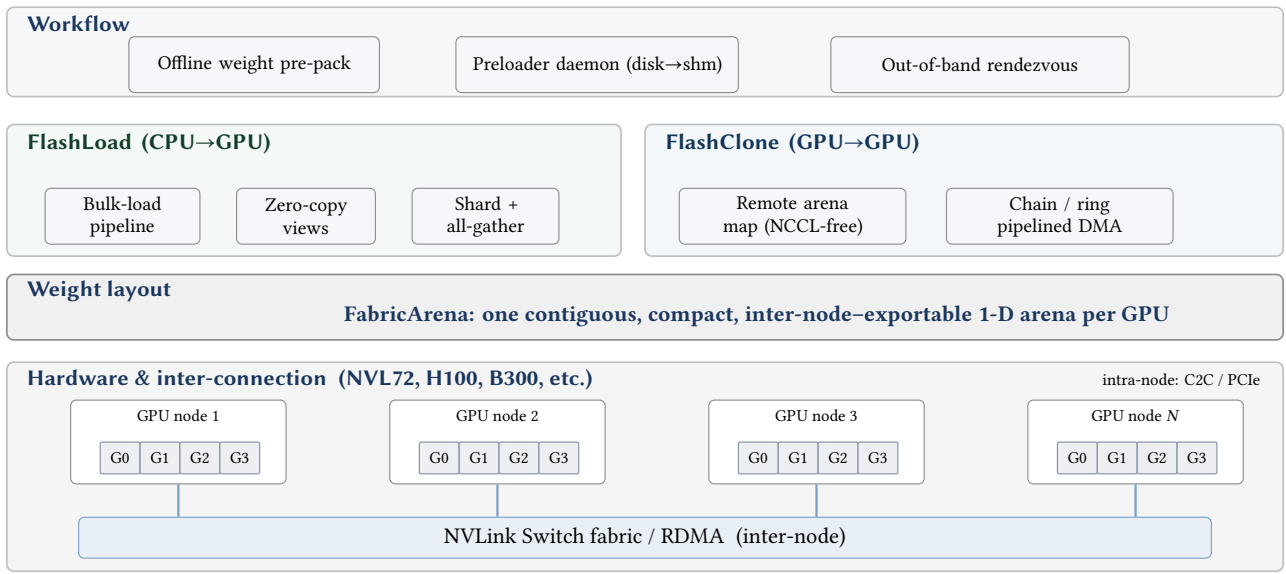
\begin{figure}[t]
\centering
\resizebox{\textwidth}{!}{%
\begin{tikzpicture}[font=\footnotesize,
  band/.style={rounded corners=2.5pt,draw=black!25,line width=0.6pt},
  sub/.style={fbbox,minimum height=7mm,inner sep=2pt,font=\fontsize{7}{8}\selectfont,align=center},
  blabel/.style={font=\scriptsize\bfseries\sffamily,color=fbnavy,anchor=west}]
\def\W{16}
\begin{scope}[shift={(0,5.95)}]
  \fill[fbnavy!4,band] (0,0) rectangle (\W,1.12);
  \node[blabel] at (0.12,0.92) {Workflow};
  \node[fbnav,sub,text width=3.0cm] at (3.1,0.4) {Offline weight pre-pack};
  \node[fbnav,sub,text width=3.4cm] at (7.4,0.4) {Preloader daemon (disk$\to$shm)};
  \node[fbnav,sub,text width=3.6cm] at (12.2,0.4) {Out-of-band rendezvous};
\end{scope}
\begin{scope}[shift={(0,3.95)}]
  \fill[fbgreen!5,band] (0,0) rectangle (7.7,1.65);
  \node[blabel,color=fbgreen!55!black] at (0.12,1.42) {FlashLoad\, (CPU$\to$GPU)};
  \node[fbgrn,sub,text width=1.8cm] at (1.45,0.5) {Bulk-load pipeline};
  \node[fbgrn,sub,text width=1.6cm] at (3.75,0.5) {Zero-copy views};
  \node[fbgrn,sub,text width=1.7cm] at (6.0,0.5) {Shard + all-gather};
  \fill[fbblue!5,band] (8.0,0) rectangle (\W,1.65);
  \node[blabel,color=fbblue!55!black] at (8.12,1.42) {FlashClone\, (GPU$\to$GPU)};
  \node[fbblb,sub,text width=2.7cm] at (9.85,0.5) {Remote arena map (NCCL-free)};
  \node[fbblb,sub,text width=2.7cm] at (13.2,0.5) {Chain / ring pipelined DMA};
\end{scope}
\begin{scope}[shift={(0,2.9)}]
  \fill[black!6,band,draw=black!45] (0,0) rectangle (\W,0.82);
  \node[blabel,color=fbnavy] at (0.12,0.61) {Weight layout};
  \node[font=\scriptsize\bfseries,color=fbnavy] at (9.4,0.3)
    {FabricArena: one contiguous, compact, inter-node--exportable 1-D arena per GPU};
\end{scope}
\begin{scope}[shift={(0,0)}]
  \fill[black!4,band] (0,0) rectangle (\W,2.62);
  \node[blabel] at (0.12,2.4) {Hardware \& inter-connection \;(NVL72, H100, B300, etc.)};
  \foreach \i/\x in {0/0.8,1/4.8,2/8.8,3/12.8}{
    \node[fbbox,fill=white,minimum width=2.7cm,minimum height=1.05cm] (nd\i) at (\x+1.35,1.62) {};
    \node[font=\tiny,color=black] at (\x+1.35,1.95) {GPU node \ifnum\i=3 $N$\else\the\numexpr\i+1\relax\fi};
    \foreach \g in {0,1,2,3} \node[draw=fbnavy!55,fill=fbnavy!8,rounded corners=0.6pt,
      minimum width=0.42cm,minimum height=0.28cm,font=\tiny] at (\x+0.62+\g*0.5,1.42) {G\g};
  }
  \node[draw=fbblue!60,fill=fbblue!10,rounded corners=2pt,minimum width=14.2cm,minimum height=0.5cm,
    font=\scriptsize,align=center] (fab) at (\W/2,0.42) {NVLink Switch fabric / RDMA \;(inter-node)};
  \foreach \i in {0,1,2,3} \draw[draw=fbblue!60,line width=0.8pt] (nd\i.south) -- (nd\i.south |- fab.north);
  \node[font=\tiny,color=black,anchor=east] at (\W-0.15,2.4) {intra-node: C2C / PCIe};
\end{scope}
\end{tikzpicture}}
\caption{\textbf{\fb{} architecture.} From the bottom: rack hardware; the
\fa{} weight-memory layout; the \fl{} (CPU$\to$GPU) and \fc{} (GPU$\to$GPU) logical
modules; and an orchestration layer. \fa{} is the waist: both modules move weights
as one contiguous byte range.}
\label{fig:arch}
\end{figure}

Figure~\ref{fig:arch} shows how the pieces compose. At the bottom is the rack:
GPU nodes connected intra-node by Grace$\leftrightarrow$GPU C2C (and PCIe) and
inter-node by the NVLink Switch fabric (or RDMA). Directly above the hardware sits
the one structural change \fb{} introduces: \fa, a contiguous, compact, and
inter-node--exportable weight memory layout (\S\ref{sec:arena}). Two logical
modules build on it: \fl{} fills an arena from CPU (bulk-load pipeline, zero-copy
views, and a shard\,+\,all-gather variant for concurrent multi-node CPU loads),
and \fc{} replicates a resident arena across nodes (NCCL-free remote mapping plus
a chain/ring pipelined DMA). An orchestration layer on top supplies the offline
pre-pack, the preloader daemon, and the out-of-band rendezvous that ferries
handles and provides the start barrier. The arena is the waist of the design: both
modules see weights as one byte range, so neither contains any per-tensor or
expert/non-expert special casing.

\subsection{FabricArena: a contiguous, inter-node--addressable layout}\label{sec:arena}
\textbf{Motivation.} C1 (\S\ref{sec:c1}) says the root problem is that weights are
scattered across many small allocations. The fix is to make them contiguous
without disturbing the engine's compute and runtime logic, which expects to
find each parameter at its own tensor handle.

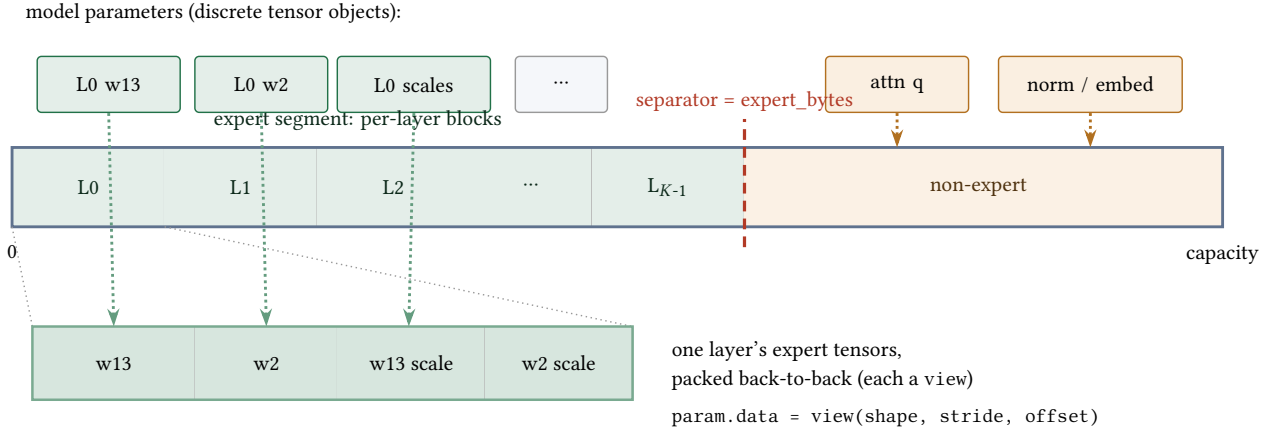
\begin{figure}[t]
\centering
\resizebox{\textwidth}{!}{%
\begin{tikzpicture}[font=\scriptsize,
  t/.style={draw=black!45,rounded corners=1.2pt,fill=fbboxbg,font=\tiny,minimum height=5.5mm,align=center,inner sep=1.5pt},
  ex/.style={t,draw=fbgreen,fill=fbgreen!12},
  ne/.style={t,draw=fbamber!80!black,fill=fbamber!14}]
\node[font=\tiny,color=black,anchor=west] at (0,4.25) {model parameters (discrete tensor objects):};
\node[ex,text width=1.3cm] (p0) at (0.95,3.55) {L0 w13};
\node[ex,text width=1.2cm] (p1) at (2.45,3.55) {L0 w2};
\node[ex,text width=1.4cm] (p2) at (3.95,3.55) {L0 scales};
\node[t,draw=black!40,text width=0.8cm] at (5.4,3.55) {$\cdots$};
\node[ne,text width=1.3cm] (p3) at (8.7,3.55) {attn q};
\node[ne,text width=1.7cm] (p4) at (10.6,3.55) {norm / embed};
\def\ay{2.15}\def\ah{0.78}
\foreach \xa/\xb/\lab in {0/1.5/L0,1.5/3.0/L1,3.0/4.5/L2,4.5/5.7/$\cdots$,5.7/7.2/{L$_{K\text{-}1}$}}{
  \fill[fbgreen!12] (\xa,\ay) rectangle (\xb,\ay+\ah);
  \draw[black!30,line width=0.4pt] (\xb,\ay) -- (\xb,\ay+\ah);
  \node[font=\tiny,color=fbgreen!35!black] at ({(\xa+\xb)/2},\ay+\ah/2) {\lab};
}
\fill[fbamber!12] (7.2,\ay) rectangle (11.9,\ay+\ah);
\node[font=\tiny,color=fbamber!35!black] at (9.5,\ay+\ah/2) {non-expert};
\draw[line width=0.8pt,draw=fbnavy!70] (0,\ay) rectangle (11.9,\ay+\ah);
\draw[densely dashed,fbred,line width=0.9pt] (7.2,\ay-0.2) -- (7.2,\ay+\ah+0.28);
\node[font=\tiny,color=fbred,anchor=south] at (7.2,\ay+\ah+0.22) {separator = expert\_bytes};
\node[font=\tiny,color=fbgreen!35!black,anchor=south] at (3.4,\ay+\ah+0.04) {expert segment: per-layer blocks};
\node[font=\tiny,color=black,anchor=north] at (0,\ay-0.04) {0};
\node[font=\tiny,color=black,anchor=north] at (11.9,\ay-0.04) {capacity};
\def\zy{0.45}\def\zh{0.72}
\draw[black!40,densely dotted] (0,\ay) -- (0.2,\zy+\zh);
\draw[black!40,densely dotted] (1.5,\ay) -- (6.1,\zy+\zh);
\foreach \xa/\xb/\lab in {0.2/1.8/w13,1.8/3.2/w2,3.2/4.65/{w13 scale},4.65/6.1/{w2 scale}}{
  \fill[fbgreen!18] (\xa,\zy) rectangle (\xb,\zy+\zh);
  \draw[black!30,line width=0.4pt] (\xb,\zy)--(\xb,\zy+\zh);
  \node[font=\tiny,color=black] at ({(\xa+\xb)/2},\zy+\zh/2) {\lab};
}
\draw[line width=0.7pt,draw=fbgreen!60] (0.2,\zy) rectangle (6.1,\zy+\zh);
\node[font=\tiny,color=black,anchor=west] at (6.35,\zy+\zh/2+0.12) {one layer's expert tensors,};
\node[font=\tiny,color=black,anchor=west] at (6.35,\zy+\zh/2-0.16) {packed back-to-back (each a \texttt{view})};
\draw[fbar,densely dotted,fbgreen!70] (p0.south) -- (1.0,\zy+\zh);
\draw[fbar,densely dotted,fbgreen!70] (p1.south) -- (2.5,\zy+\zh);
\draw[fbar,densely dotted,fbgreen!70] (p2.south) -- (3.9,\zy+\zh);
\draw[fbar,densely dotted,fbamber!80!black] (p3.south) -- (8.7,\ay+\ah);
\draw[fbar,densely dotted,fbamber!80!black] (p4.south) -- (10.6,\ay+\ah);
\node[font=\tiny,color=black,anchor=west] at (6.35,\zy-0.18) {\texttt{param.data = view(shape, stride, offset)}};
\end{tikzpicture}}
\caption{\textbf{\fa{} layout and zero-copy views.} Discrete model parameters
(top) are not given private storage; each parameter's data is an
\code{as\_strided(shape, stride, offset)} view onto a byte range of one
contiguous arena (bottom), laid out $[\,\text{experts}\mid\text{non-experts}\,]$.
The weights exist exactly once and are served in place; both \fl{} (host$\to$GPU)
and \fc{} (GPU$\to$GPU) only ever fill the flat buffer.}
\label{fig:arena}
\end{figure}

\textbf{Mechanism.} \fa{} resolves this tension with the CUDA virtual-memory
management (VMM) API. We reserve one large, contiguous virtual address
range per GPU and back it with a single physical allocation (\code{cuMemCreate}
$\to$ \code{cuMemAddressReserve} $\to$ \code{cuMemMap} $\to$ \code{cuMemSetAccess}).
Every weight tensor then lives at a distinct offset inside this one arena.
The model's parameters keep their familiar tensor interface, but their storage is
now a slice of a single 1-D buffer (Figure~\ref{fig:arena}). Crucially, the
allocation is created with a fabric handle type
(\code{CU\_MEM\_HANDLE\_TYPE\_FABRIC}), which is what lets a \mbox{64-byte} handle
for the whole arena be exported and imported by a GPU on another node within the
same fabric/IMEX domain, making the layout not just contiguous but
inter-node addressable (used by \fc, \S\ref{sec:remotemap}). The arena
capacity is rounded up to the VMM allocation granularity, and seed and clone apply
the same rounding, so both sides always agree on the exact byte count.

\textbf{Layout.} We lay the arena out as two contiguous regions,
$[\,\text{experts}\mid\text{non-experts}\,]$, with the separator offset equal to
the total expert byte size (Figure~\ref{fig:arena}). This mirrors how MoE engines
already group weights and keeps the two natural populations (the many large
expert tensors and the comparatively few attention/embedding/norm tensors) each
in one run. The on-disk pre-packed image (\S\ref{sec:workflow}) is byte-identical
to this in-GPU order, so filling the arena is a pure copy with no parse-and-reshape
on the critical path. Because the byte layout already matches what the serving
kernels expect, the only special case the engine needs is the boundary between the
two regions; everything else is offset arithmetic.

\subsection{Zero-copy serving and a uniform transport}\label{sec:zerocopy}
\textbf{Principle.} Once the weights are one contiguous image, we never want a
second copy of them anywhere. We bind each parameter's data pointer to an
\code{as\_strided} view into the arena at the offset, shape, and stride
recorded for it (Figure~\ref{fig:arena}). Reading a weight reads arena bytes in
place: there is exactly one copy in GPU memory ($1\times$ occupancy), and the
per-tensor views are re-derived from the offset table on each side rather than
materialized.

\textbf{Why it unifies the transport.} Because the destination is always ``the
arena at offset $o$,'' the source can be anything that speaks bytes: a
host pinned buffer (the CPU-load path) or a remote GPU's mapped arena (the
cross-node path). Both reduce to filling one contiguous 1-D range, so a single
mover, a copy-engine DMA, serves every case. The copy engine is the right
primitive here for three reasons that Table~\ref{tab:transport} makes precise: it
is genuinely zero-copy into the final layout, it uses no streaming
multiprocessors (leaving the SMs free, unlike NCCL's kernel-based collectives),
and it reaches the highest peak bandwidth we measure on the fabric
($\sim$$826$\gbs, versus $\sim$$651$\gbs{} for NCCL on the same link,
\S\ref{sec:micro}).

\begin{table}[t]
\centering\small
\caption{Transport comparison for weight movement. \fb{} uses copy-engine DMA,
which is zero-copy into the final layout, occupies no SMs, moves the image as one
bulk transfer, works both intra- and inter-node, and reaches the highest measured
peak. Marks: \textcolor{fbgreen}{\ding{51}}~supported, \textcolor{fbamber!85!black}{partial},
\textcolor{fbred}{\ding{55}}~not supported.}
\label{tab:transport}
\begin{tabular}{@{}l c c c c c@{}}
\toprule
& Zero-copy & Single bulk & SM-free & Cross-node & Peak BW \\
& ($1\times$ mem) & transfer & (copy engine) & (NVLink/RDMA) & (measured) \\
\midrule
SafeTensors        & \textcolor{fbred}{\ding{55}} & \textcolor{fbred}{\ding{55}} & \textcolor{fbred}{\ding{55}} & ---       & low \\
InstantTensor      & \textcolor{fbamber!85!black}{partial} & \textcolor{fbamber!85!black}{partial} & \textcolor{fbred}{\ding{55}} & ---       & low--mid \\
Tensor R-Fork (NCCL) & \textcolor{fbred}{\ding{55}} & \textcolor{fbred}{\ding{55}} & \textcolor{fbred}{\ding{55}} & \textcolor{fbgreen}{\ding{51}} & $\sim$651\gbs \\
\textbf{\fb{} (DMA)} & \textcolor{fbgreen}{\ding{51}} & \textcolor{fbgreen}{\ding{51}} & \textcolor{fbgreen}{\ding{51}} & \textcolor{fbgreen}{\ding{51}} & $\sim$\textbf{826}\gbs \\
\bottomrule
\end{tabular}
\end{table}

\subsection{Bulk-load pipeline}\label{sec:bulkpipe}
\begin{figure}[t]
\centering
\begin{tikzpicture}[font=\scriptsize, b/.style={fbbox,minimum height=8mm,align=center,inner sep=2.5pt,font=\scriptsize}]
\node[fblbl,color=fbred,anchor=east] at (-0.35,1.8) {daemon};
\node[fbredb,b,text width=1.25cm] (disk) at (0.85,1.8) {disk (NVMe)};
\node[fbredb,b,text width=1.9cm] (shm) at (5.0,1.8) {NUMA-pinned shm};
\draw[fbar,fbred] (disk)--node[fblbl,above,yshift=1pt]{O\_DIRECT read}(shm);
\node[fblbl,color=fbblue,anchor=east] at (-0.35,0) {engine};
\node[fbblb,b,text width=1.9cm] (start) at (1.0,0) {engine startup};
\node[b,text width=1.7cm] (pin) at (4.7,0) {PIN (register)};
\node[fbgrn,b,text width=2.2cm] (arena) at (8.1,0) {H2D $\to$ \fa};
\node[b,text width=1.5cm] (serve) at (11.0,0) {serve (views)};
\draw[fbar] (start)--(pin); \draw[fbar](pin)--(arena); \draw[fbar](arena)--(serve);
\draw[fbar,fbred,densely dashed] (shm.south)--node[fblbl,right,xshift=1pt]{\texttt{.ready} (io\_wait$\approx$0)}(pin.north);
\draw[decorate,decoration={brace,amplitude=4pt,mirror},color=black] (pin.south west)++(0,-0.18) -- ++(6.8,0)
  node[midway,below,fblbl]{critical path $=$ PIN $+$ H2D only};
\end{tikzpicture}
\caption{\textbf{Bulk-load pipeline.} The preloader daemon stages disk$\to$pinned
shm concurrently with engine startup; the engine then only PINs and H2Ds the
already-staged image into the arena. The slow cold read is hidden
(\code{io\_wait}$\approx 0$), so the engine's critical path is pure data movement.}
\label{fig:loadpipe}
\end{figure}
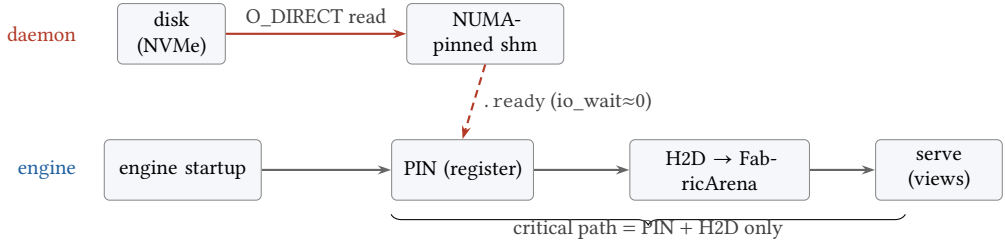

\textbf{From contiguity to bandwidth.} With the arena in place, the host$\to$GPU
load is a single contiguous H2D copy, which already saturates the C2C link at the
measured $\sim$$185$\gbs. The remaining cost on a cold boot is the slow disk read
that feeds that copy. We remove it from the critical path with a preloader daemon
(Figure~\ref{fig:loadpipe}): a separate process reads each rank's pre-packed files
\code{O\_DIRECT} from disk into NUMA-pinned shared memory ahead of and
concurrently with engine startup. When the engine reaches the fill point it finds
the pages already staged (signalled by a \code{.ready} flag, so the wait
\code{io\_wait}$\approx 0$), registers them for DMA (PIN), and H2Ds them straight
into the arena. The engine-visible critical path is thus only PIN$+$H2D; the cold
read, which dominates wall-clock, overlaps work the engine had to do anyway. Pinned
buffers are NUMA-bound to the Grace socket nearest each GPU so the H2D does not
cross sockets, and the H2D itself runs as a few large chunks over multiple streams
to keep the copy engine busy.

\subsection{Concurrent multi-node CPU load: shard and all-gather}\label{sec:shard}
\textbf{Motivation.} The single-node path still has one node read the entire
checkpoint, and that cold read dominates its cost. When the same model comes up on
$N$ nodes at once (e.g.,\ $N$ data-parallel serving replicas), we can do better by
using the rack's aggregate disk and CPU bandwidth.

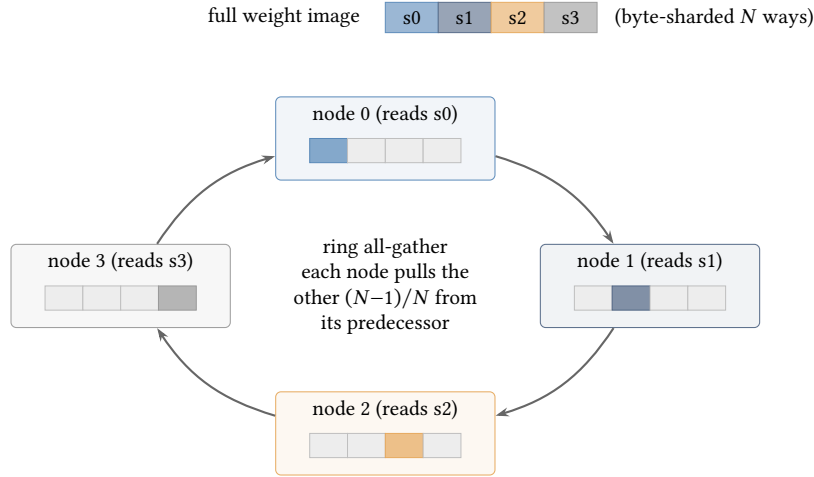
\begin{figure}[t]
\centering
\begin{tikzpicture}[font=\scriptsize]
\node[font=\scriptsize,color=black,anchor=east] at (-0.2,3.55) {full weight image};
\foreach \i/\c in {0/fbblue,1/fbnavy,2/fbamber,3/fbgray}{
  \fill[\c!45,draw=\c!70,line width=0.4pt] (\i*0.7,3.35) rectangle (\i*0.7+0.7,3.75);
  \node[font=\scriptsize,color=black] at (\i*0.7+0.35,3.55) {s\i};
}
\node[font=\scriptsize,color=black,anchor=west] at (4*0.7+0.1,3.55) {(byte-sharded $N$ ways)};
\foreach \n/\px/\py/\c in {0/0/1.95/fbblue, 1/3.5/0/fbnavy, 2/0/-1.95/fbamber, 3/-3.5/0/fbgray}{
  \node[fbbox,draw=\c!70,fill=\c!6,minimum width=2.9cm,minimum height=1.1cm] (nd\n) at (\px,\py) {};
  \node[font=\scriptsize,color=black] at (\px,\py+0.32) {node \n\ (reads s\n)};
  \foreach \i/\cc in {0/fbblue,1/fbnavy,2/fbamber,3/fbgray}{
    \ifnum\i=\n
      \fill[\cc!55,draw=\cc!75,line width=0.3pt] (\px-1.0+\i*0.5,\py-0.32) rectangle (\px-1.0+\i*0.5+0.5,\py+0.0);
    \else
      \fill[black!7,draw=black!25,line width=0.3pt] (\px-1.0+\i*0.5,\py-0.32) rectangle (\px-1.0+\i*0.5+0.5,\py+0.0);
    \fi
  }
}
\draw[fbar] (nd0) to[bend left=20] (nd1);
\draw[fbar] (nd1) to[bend left=20] (nd2);
\draw[fbar] (nd2) to[bend left=20] (nd3);
\draw[fbar] (nd3) to[bend left=20] (nd0);
\node[font=\scriptsize,color=black,align=center] at (0,0) {ring all-gather\\each node pulls the\\other $(N{-}1)/N$ from\\its predecessor};
\end{tikzpicture}
\caption{\textbf{Shard\,+\,fabric all-gather for concurrent multi-node CPU load.}
$N$ nodes split the cold disk read $N$ ways ($\sim 1/N$), then reassemble the full
image with one NCCL-free ring all-gather over NVLink.}
\label{fig:shard}
\end{figure}

\textbf{Mechanism.} The checkpoint is sharded by bytes across the $N$ nodes: each
reads only its $1/N$ slice from its own NVMe (in parallel with the others) and
H2Ds it into the correct offset of its full-size arena. Each node now holds only a
$1/N$ fragment, so the nodes all-gather the missing pieces over the NVLink
fabric, reusing the same exportable arena handles, with no NCCL group
(Figure~\ref{fig:shard}). We run the gather as a ring: in each round a node
reads only from its predecessor over an independent link, so every GPU has exactly
one reader and per-link bandwidth stays high and uniform however large $N$ grows
(principle P4). The dominant cold read therefore scales as $\sim 1/N$, while the
gather adds only a fixed, bandwidth-bound pass over the missing $(N{-}1)/N$ at the
$\sim$$740$\gbs{} fabric rate.

\subsection{Remote memory mapping: replacing the communicator}\label{sec:remotemap}
\textbf{Motivation.} C2 (\S\ref{sec:c2}) showed that NCCL's communicator standup,
not the transfer, dominates cross-node replication. The fabric's per-GPU bandwidth
is so high that the communicator is the bottleneck. Patching NCCL to remove the
standup is conceivable but invasive and hard to maintain; instead we sidestep
collectives entirely.

\begin{figure}[t]
\centering
\begin{tikzpicture}[font=\small, b/.style={fbbox,minimum height=7mm,align=center,inner sep=3pt,font=\footnotesize,text width=3.4cm}]
\node[font=\footnotesize\bfseries,color=fbblue] at (0,3.7) {Seed};
\node[fbblb,b] (s1) at (0,3.0) {weights $\to$ \fa};
\node[fbblb,b] (s2) at (0,2.0) {export 64-byte handle};
\node[fbblb,b] (s3) at (0,1.0) {publish manifest};
\node[fbblb,b,fill=fbblue!4] (s4) at (0,0.0) {\textbf{Passive} (holds arena)};
\node[font=\footnotesize\bfseries,color=fbgreen!60!black] at (7.8,3.7) {Clone};
\node[fbgrn,b] (c1) at (7.8,3.0) {fetch manifest (out-of-band)};
\node[fbgrn,b] (c2) at (7.8,2.0) {import handle $+$ map remote ($\sim$10\,ms, IMEX)};
\node[fbgrn,b] (c3) at (7.8,1.0) {grant self read access};
\node[fbgrn,b] (c4) at (7.8,0.0) {DMA read $\to$ rebind views};
\draw[fbar,fbblue](s1)--(s2);\draw[fbar,fbblue](s2)--(s3);\draw[fbar,fbblue](s3)--(s4);
\draw[fbar,fbgreen](c1)--(c2);\draw[fbar,fbgreen](c2)--(c3);\draw[fbar,fbgreen](c3)--(c4);
\draw[fbar,densely dashed,black!55] (s3.east)--node[fblbl,above,align=center]{manifest\\(few hundred B, once)}(c1.west);
\draw[{Stealth[length=6pt]}-,line width=1.3pt,fbamber!70!black] (s4.east)--(c4.west)
  node[midway,above=0pt,fblbl,color=fbamber!40!black,align=center]{NVLink DMA\\(the only data movement)};
\end{tikzpicture}
\caption{\textbf{Remote arena mapping (NCCL-free).} The seed exports one 64-byte
fabric handle and goes passive; the clone imports it, maps the seed's arena, grants
itself read access ($\sim$10\,ms), and pulls the bytes with a DMA over NVLink, with no
process group, no runtime rendezvous. Only the tiny manifest crosses nodes out of
band.}
\label{fig:remotemap}
\end{figure}
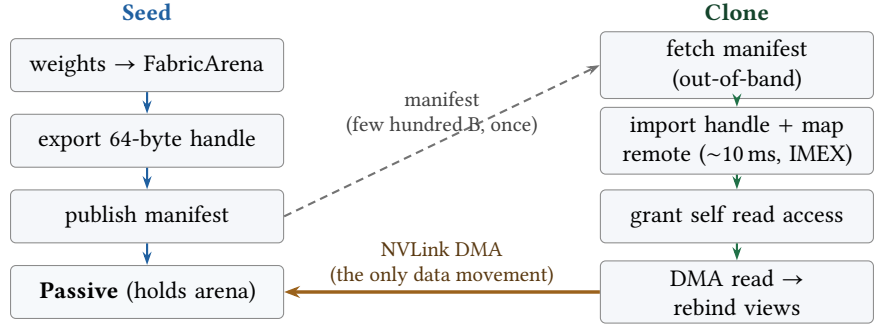

\textbf{Mechanism.} We treat the rack as one shared address space (a partitioned global address space, or PGAS):
because \fa{} memory is created with a fabric handle (\S\ref{sec:arena}), a seed can
export a 64-byte handle to its whole arena, and any other node can
import that handle and map the seed's physical memory into its own virtual
address space, granting itself read access (gated by the IMEX security
service). The seed is never in the authorization loop; it exports once and goes
passive. After mapping, a clone reads the seed's weights with a plain DMA over
NVLink, straight into its own arena (Figure~\ref{fig:remotemap}). The only
per-event cost is the export/import, measured at $\sim$$2$--$15$\,ms per handle and
independent of payload size; the handful of handles are ferried by an out-of-band
TCP rendezvous (\S\ref{sec:workflow}) that can overlap the rest of startup. The
result is the $\sim$$10$\,ms setup that replaces NCCL's $10$--$110$\,s
(Figure~\ref{fig:nccl}).

\textbf{Transports: NVLink (IMEX) and RDMA.} The fabric-handle export/import we use
here is the NVLink path: it relies on IMEX, the internode memory-sharing service
specific to multi-node NVLink (MNNVL) platforms such as NVL72 and DGX/HGX
GB200/GB300~\cite{imex}. The mechanism is otherwise transport-agnostic, needing
only a primitive that grants one node direct read access to another's memory plus a
small out-of-band channel for the access token; on clusters without MNNVL that role
falls to \textbf{RDMA} memory registration and one-sided reads, which we discuss in
\S\ref{sec:beyond}. We evaluate the IMEX/NVLink implementation here and leave the
RDMA backend to future work.

\subsection{Star vs.\ chain: scaling the broadcast}\label{sec:topo}
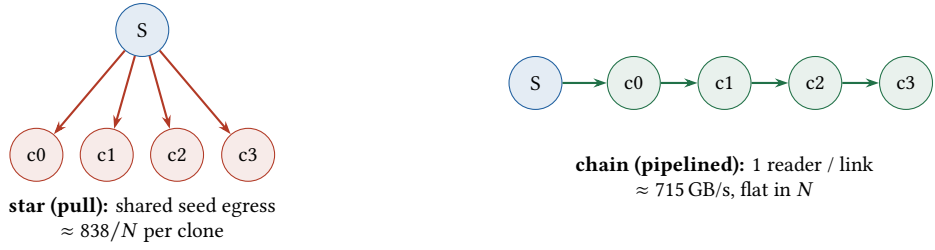
\begin{figure}[t]
\centering
\begin{tikzpicture}[font=\small, g/.style={circle,draw,minimum size=7mm,font=\scriptsize,inner sep=0pt}]
\node[g,fill=fbblue!12,draw=fbblue] (ss) at (1.4,1.25) {S};
\foreach \i/\x in {0/0,1/0.93,2/1.86,3/2.8} \node[g,fill=fbred!10,draw=fbred] (sc\i) at (\x,-0.4) {c\i};
\foreach \i in {0,1,2,3} \draw[fbar,fbred] (ss)--(sc\i);
\node[font=\scriptsize,align=center] at (1.4,-1.25) {\textbf{star (pull):} shared seed egress\\$\approx 838/N$ per clone};
\begin{scope}[shift={(7.2,0)}]
\node[g,fill=fbblue!12,draw=fbblue] (cs) at (-0.6,0.55) {S};
\foreach \i/\x in {0/0.7,1/1.9,2/3.1,3/4.3} \node[g,fill=fbgreen!10,draw=fbgreen] (cc\i) at (\x,0.55) {c\i};
\draw[fbar,fbgreen] (cs)--(cc0);\draw[fbar,fbgreen](cc0)--(cc1);\draw[fbar,fbgreen](cc1)--(cc2);\draw[fbar,fbgreen](cc2)--(cc3);
\node[font=\scriptsize,align=center] at (1.9,-0.7) {\textbf{chain (pipelined):} 1 reader / link\\$\approx 715$\,GB/s, flat in $N$};
\end{scope}
\end{tikzpicture}
\caption{\textbf{Broadcast topologies.} Star: every clone pulls from one
seed, sharing its egress ($\sim$$838/N$). Chain: each clone reads its
predecessor over an independent link, chunk-pipelined, so per-clone bandwidth is
flat in $N$. The seed is passive in both.}
\label{fig:topo}
\end{figure}

Remote mapping makes one clone fast; bringing up many requires care
(C3, \S\ref{sec:c3}). We designed and compared two topologies
(Figure~\ref{fig:topo}). In the \textbf{star} (pull) topology every clone
maps and reads the same seed arena; it is the simplest scheme, one copy per
clone, but all $N$ readers share the seed GPU's single NVLink egress, so per-clone
bandwidth falls as $\sim$$838/N$. In the \textbf{chain} topology the clones form a
line, $\text{seed}\to c_0 \to c_1 \to \cdots$, each reading from its
predecessor rather than the seed. Every hop is a distinct NVLink, so the
transfers run on independent links in parallel and per-clone bandwidth stays near
the single-link ceiling regardless of $N$ (principle P4). To keep every link busy,
the arena streams down the line in large fixed-size chunks, so a node receives
chunk $j$ while its predecessor pulls chunk $j{+}1$ (software pipelining over the
copy engine, synchronized by lightweight in-fabric flags, with no NCCL or SM kernels).
The only cost is a one-time pipeline-fill latency, a small head-to-tail
gradient, amortized over the chunks; chunk size is the lone knob, trading
per-chunk synchronization against fill-tail length (we analyze it in
\S\ref{sec:disc}). Chain is therefore the default whenever more than one clone is
filled; all subsequent multi-node measurements use it.

\subsection{Workflow: pay the heavy work once}\label{sec:workflow}
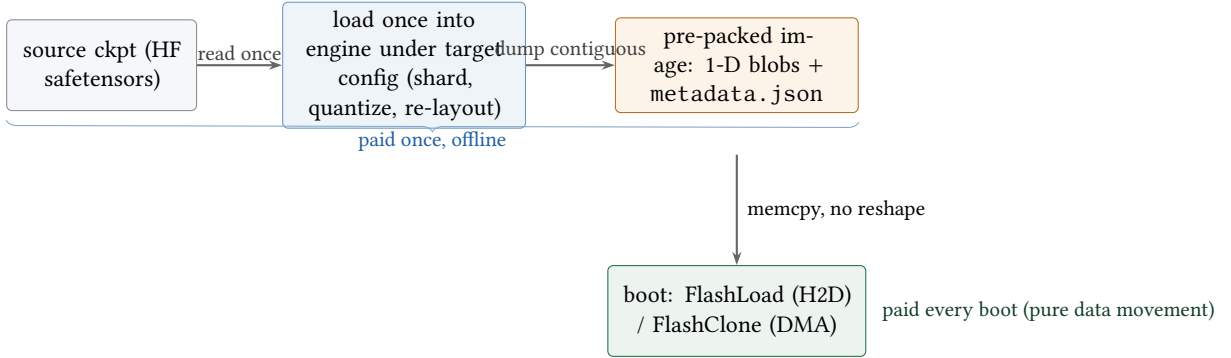
\begin{figure}[t]
\centering
\begin{tikzpicture}[font=\small,
  s/.style={fbbox,align=center,inner sep=3pt,minimum height=12mm,font=\footnotesize},
  sg/.style={s,draw=fbblue!65,fill=fbblue!7},
  ck/.style={s,draw=fbamber!80!black,fill=fbamber!10},
  bt/.style={s,draw=fbgreen,fill=fbgreen!8}]
\node[s,text width=2.3cm] (src) at (0,1.9) {source ckpt (HF safetensors)};
\node[sg,text width=3.0cm] (sg) at (4.0,1.9) {load once into engine under target config (shard, quantize, re-layout)};
\node[ck,text width=3.0cm] (img) at (8.4,1.9) {pre-packed image: 1-D blobs $+$ \code{metadata.json}};
\draw[fbar] (src)--node[fblbl,above]{read once}(sg);
\draw[fbar] (sg)--node[fblbl,above]{dump contiguous}(img);
\draw[decorate,decoration={brace,mirror,amplitude=4pt},draw=fbblue!55]
  (src.south west)++(0,-0.12) -- ($(img.south east)+(0,-0.12)$)
  node[midway,below,font=\scriptsize,color=fbblue]{paid once, offline};
\node[bt,text width=3.2cm] (boot) at (8.4,-1.35) {boot: \fl{} (H2D) / \fc{} (DMA)};
\draw[fbar] (img.south)++(0,-0.55) -- (boot.north)
  node[midway,right,font=\scriptsize,color=black]{memcpy, no reshape};
\node[font=\scriptsize,color=fbgreen!55!black,anchor=west] at (boot.east) {\;paid every boot (pure data movement)};
\end{tikzpicture}
\caption{\textbf{Workflow.} A one-off offline pre-pack puts the checkpoint in its
final, distribution-baked, byte-for-byte arena order. Every boot thereafter, whether a CPU
load or cross-node clone, is a pure copy with \code{as\_strided} views and no
per-parameter reshape.}
\label{fig:workflow}
\end{figure}

The last principle (P5) is realized by the orchestration layer
(Figure~\ref{fig:workflow}). All distribution-dependent work (tensor-parallel
sharding, fp8/mxfp4 quantization, and any kernel-required re-layout) is done
once, offline, by a pre-pack step: the source checkpoint is loaded
once into the engine under the exact target serving configuration, and each
rank's now-resident weights are serialized to disk as contiguous byte ranges
(per-layer expert blobs followed by a single non-expert blob) plus a small
\code{metadata.json} that records each parameter's name $\to$ (offset, shape,
stride, dtype). Because this on-disk order is byte-identical to the in-GPU arena
layout (\S\ref{sec:arena}), every subsequent boot is pure data movement: there is
no per-parameter dequantization, permutation, or reshape on the critical path. The
remaining orchestration is deliberately thin: the preloader daemon
(\S\ref{sec:bulkpipe}) and a small out-of-band TCP rendezvous that exchanges
fabric handles and provides a start barrier for the chain/all-gather, a single
sub-millisecond round-trip in place of any communicator.

\section{Implementation}\label{sec:impl}

We implemented \fb{} as a lightweight, self-contained native library plus a thin
integration into the SGLang inference engine (a DeepSeek-V4--capable development
build; CUDA~13.0, \code{torch 2.9.1+cu130}, NCCL~2.27.7). The library is
deliberately credential-free: it manipulates fabric memory and reads/writes local
manifest files only; all cluster orchestration (node selection, manifest delivery,
the rendezvous server) lives in separate scripts. The engine-side integration is a
single patch of roughly $5{,}800$ added and $280$ removed lines across $19$ files;
Table~\ref{tab:impl} summarizes the integration surface.

\begin{table}[t]
\centering\small
\caption{Integration surface. The \fb{} library is standalone; the engine changes
are confined to the loader and weight-handling paths.}
\label{tab:impl}
\begin{tabular}{@{}l >{\raggedright\arraybackslash}p{10.2cm}@{}}
\toprule
Component & Contents \\
\midrule
\fb{} native core & arena (VMM), importer, pluggable allocator, ring/chain movers,
  copy backends (copy-engine / TMA / SM), IMEX preflight, manifest, pinned-shm
  stager, host mover; one stable C ABI \\
\fb{} Python & seed / clone / DP loaders, transport, rendezvous, chunk policy,
  preflight (no torch needed for orchestration helpers) \\
SGLang patch & $\sim$5.8k added / $\sim$0.3k removed lines across 19 files,
  concentrated in the model loader, MoE/linear layers, the mxfp4 quantizer, weight
  utilities, and the remote-instance connector \\
\bottomrule
\end{tabular}
\end{table}

\paragraph{Arena and the engine's tensors.}
The engine's loader already meta-initializes the model (parameters with no backing
storage) and then materializes weights. We hook the materialization step: instead
of allocating per-parameter storage, the loader allocates one unified \fa{} per GPU
and rebinds every \code{param.data} to an \code{as\_strided} view into it
(\S\ref{sec:zerocopy}); the rest of the engine is unchanged and never learns there
is a single allocation underneath. In practice we distinguish two weight
populations because they are produced differently: expert weights
(MoE \code{w13}/\code{w2} and their quantization scales) are written in their
post-quantization, fused/shuffled form so the load can scatter them
directly, while non-expert weights (attention, embedding, norms) keep the
checkpoint's native layout and flow through the engine's ordinary TP-sharding path.
The arena's $[\,\text{experts}\mid\text{non-experts}\,]$ split (\S\ref{sec:arena})
falls out of exactly this distinction; the separator is the total expert byte size,
which for both evaluation models is $4096$-byte aligned, so the non-expert views
inherit a valid dtype alignment without padding.

\paragraph{Pre-pack, model, and parallelism coverage.}
The offline pre-pack (\S\ref{sec:workflow}) reuses the engine's own
\code{weight\_loader} and \code{process\_weights\_after\_loading}, so the produced
bytes are identical to what the engine would hold at serving time; there is no
separate, drift-prone serializer. Because the engine's distributed sharding is a
local operation keyed off the parallel rank/size captured at module init (no
collective runs on the load path), we produce all target ranks on a single
physical GPU by faking the (rank, size) context per rank, which makes pre-pack a
single-GPU offline job even when a full per-rank shard is large.

Two layers should be kept separate when reasoning about coverage. The \fa{}
layout, the zero-copy views, and the entire \fc{} replication path treat the
weights as one opaque contiguous byte image plus an offset table, so they are
model- and parallelism-agnostic by construction: any image a seed produces can be
mapped and cloned with no per-tensor or per-architecture logic. Coverage is scoped
only by the \emph{producer} of that image, the \fl{} fast path. Our prototype
gates the fast loader to \code{DeepseekV4ForCausalLM} and bakes in the
DeepSeek-V4 MoE conventions (gate/up\,$\to$\,\code{w13} fusion, the \code{w2}
down-projection, fp8 block scales, and the mxfp4 expert quantizer); the generic
per-shard prefetch we add to SGLang's linear and fused-MoE layers, by contrast,
benefits any model. Extending the fast path to the other MoE families SGLang
serves (e.g.\ Qwen3-30B-A3B, GLM, Nemotron) is a matter of registering each
family's expert-fusion and quantization convention in the pre-pack and scatter,
not a redesign; the contiguous-arena machinery beneath is unchanged. On the
parallelism axis the fast expert scatter handles tensor parallelism (a per-shard
\code{narrow}) and expert parallelism (a global$\to$local expert remap keyed off
\code{moe\_ep\_rank}/\code{moe\_ep\_size}); data-parallel and DP-attention layouts
are captured verbatim by the pre-pack, since they only change which bytes a rank
holds; non-trivial expert-placement maps (redundant-expert EPLB) fall back to the
engine's standard loader; and pipeline parallelism, whose per-rank layer subset
the current pre-pack does not yet emit, is left to future work.

\paragraph{NUMA, pinning, and the host mover.}
The host path is tuned for the Grace--Blackwell topology. Staging buffers are
\code{mmap}'d and bound to the Grace socket nearest the target GPU (\code{mbind}),
and registered with \code{cudaHostRegister}, which, unlike \code{cudaMallocHost},
keeps the pages on local LPDDR rather than migrating them, so the H2D reads from
the near socket. The mover reads files with many threads under \code{O\_DIRECT}
(falling back to buffered I/O if a container/overlay filesystem rejects direct
reads at runtime) and issues the H2D as large multi-stream chunks. Peak host memory
is held to arena $+$ one staging buffer rather than a second full copy of
the weights, which matters because a Pro arena is already $217$\,GB per GPU; in
daemon mode the engine attaches the daemon's already-staged shm (the same physical
pages, no copy) and registers it in place.

\paragraph{Three runtime details.}
Three details proved important in practice. \textbf{(1)}~A clone must not use
the engine's native remote-instance load format, which would make the engine
handshake the seed's NCCL send-weights group and block the seed's own serving; the
fabric path dispatches on its own switch and needs only the published manifest.
\textbf{(2)}~The seed must keep its arena alive for the service lifetime (we hold a
strong reference on the served model); if the seed exits, exported handles go stale
and clones must re-fetch. \textbf{(3)}~All participants must run as the same user
sharing an IMEX channel (the fabric security model is per-user), so a one-line
preflight check is run on each node, and the system falls back to the existing NCCL
path if IMEX cannot be configured.

\section{Evaluation}\label{sec:eval}

We evaluate on the GB300 NVL72 platform of Table~\ref{tab:env}, with the two models
of Table~\ref{tab:models} at TP4. All transfer timings measure data movement only
(CUDA-event timed; for microbenchmarks we report the minimum over 30 runs); the one-time fabric map and the
start barrier are reported separately and excluded from bandwidth, as they are
setup, not per-transfer cost. All reported runs validated correct: chain
replication matched the seed $8/8$ byte-exact, and the DP loads passed byte-exact
arena verification.
We answer four questions: how the fabric transport compares to NCCL at the
primitive level (\S\ref{sec:micro}); single-node \fl{} (\S\ref{sec:e-load});
concurrent multi-node \fl{} (\S\ref{sec:e-dp}); and cross-node \fc{}
(\S\ref{sec:e-clone}). Where a quantity was measured at several dates we use the
most recent.

\subsection{Microbenchmark: fabric transport vs.\ NCCL}\label{sec:micro}
\begin{figure}[t]
\centering
\begin{tikzpicture}
\begin{groupplot}[
  group style={group size=3 by 1, horizontal sep=0.5cm,
               ylabels at=edge left, yticklabels at=edge left},
  width=0.375\textwidth, height=4.7cm, fbbase, ymin=0, ymax=880,
  ytick={0,200,400,600,800}, ylabel={GB/s},
  xmode=log, log basis x=2, title style={yshift=11pt},
  x tick label style={font=\scriptsize,rotate=30,anchor=north east},
  every axis plot/.append style={very thick,mark size=1.1pt},
]
\nextgroupplot[title={(a) 1$\to$1 broadcast},
  xtick={1,16,256,4096,32768}, xticklabels={1MB,16MB,256MB,4GB,32GB},
  legend columns=-1, legend style={at={(0.5,1.0)},anchor=south,draw=none,font=\scriptsize,
    /tikz/every even column/.append style={column sep=0.3cm}},legend cell align=left]
\addplot[fbgreen,mark=*] coordinates {(1,47.3)(2,88.7)(4,169.6)(8,269.7)(16,406.7)(32,527.5)(64,643.3)(128,719.8)(256,780.5)(512,799.8)(1024,816)(2048,823.3)(4096,826)(8192,826.2)(16384,825.5)(32768,825.7)}; \addlegendentry{FlashBoot}
\addplot[fbred,mark=square*,mark size=0.9pt] coordinates {(1,14.19)(2,29.6)(4,55.79)(8,114.19)(16,217.59)(32,332.84)(64,413.1)(128,495.28)(256,564.81)(512,607.53)(1024,630.57)(2048,642.53)(4096,647.6)(8192,650.15)(16384,650.35)(32768,651.28)}; \addlegendentry{NCCL}
\nextgroupplot[title={(b) 1$\to$N broadcast (5 nodes)},
  xtick={1,16,256,4096,32768}, xticklabels={1MB,16MB,256MB,4GB,32GB}]
\addplot[fbgreen,mark=*] coordinates {(1,48.1)(2,70.8)(4,172.9)(8,285.6)(16,408.3)(32,537.5)(64,604)(128,676)(256,702.2)(512,718.5)(1024,729.8)(2048,737)(4096,739.9)(8192,740.2)(16384,739.9)(32768,739.6)};
\addplot[fbred,mark=square*,mark size=0.9pt] coordinates {(1,10.34)(2,22.97)(4,59.32)(8,113.4)(16,214.91)(32,313)(64,384.47)(128,471.37)(256,540.51)(512,568.66)(1024,596.8)(2048,614.18)(4096,623.48)(8192,634.6)(16384,635.5)(32768,637.02)};
\nextgroupplot[title={(c) all-gather (5 nodes)},
  xtick={1,16,256,8192}, xticklabels={1MB,16MB,256MB,8GB}]
\addplot[fbgreen,mark=*] coordinates {(1,46.1)(2,97.2)(4,121.5)(8,258.9)(16,385.1)(32,518.8)(64,622.9)(128,678.8)(256,709)(512,722.2)(1024,729.2)(2048,737.8)(4096,740)(8192,739.7)};
\addplot[fbred,mark=square*,mark size=0.9pt] coordinates {(1,22.26)(2,42.78)(4,89.35)(8,166.11)(16,252.93)(32,324.44)(64,357)(128,403.13)(256,441.5)(512,462.32)(1024,479.64)(2048,489.49)(4096,494.47)(8192,497.19)};
\end{groupplot}
\end{tikzpicture}
\caption{\textbf{Fabric copy-engine vs.\ NCCL, cross-node bandwidth} (5 GPUs, 1
per node; per-link for broadcast, bus-BW for all-gather; critical path = slowest
participant; broadcasts swept to 32\,GB, all-gather to its 8\,GB per-node shard).
\fb{} leads at every size and every collective. The gap is largest for small
(latency-bound) messages and for all-gather.}
\label{fig:micro}
\end{figure}
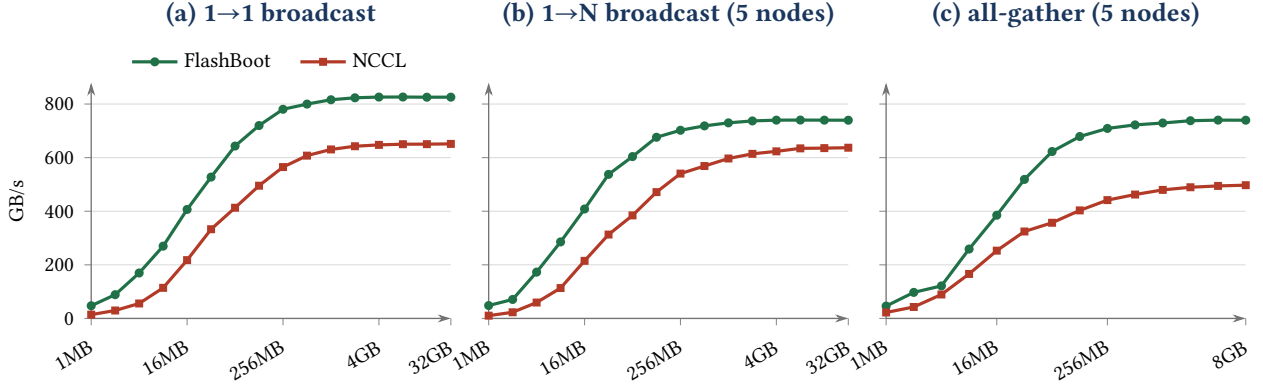

We first isolate the transport from the loader. On a single node, the contiguous
H2D over C2C sustains $\sim$$185$\gbs{} per GPU, near the $223.5$\gbs{} C2C ceiling
and limited only by it, not by software (it does not inflate when ranks are
desynchronized, confirming it is the true contended rate). The more interesting
comparison is cross-node, where \fb's copy-engine DMA replaces NCCL collectives.
Figure~\ref{fig:micro} sweeps payload from 1\,MB to 32\,GB (to 8\,GB for
all-gather) over 5 cross-node GPUs for three collectives; Table~\ref{tab:micro}
summarizes the speedups.

\begin{table}[t]
\centering\small
\caption{Fabric/NCCL bandwidth speedup. Small messages are latency-bound (fabric
\code{copy\_d2d} $\sim$$22\,\mu$s/op vs.\ NCCL $69$--$176\,\mu$s/op); large messages
are bandwidth-bound, where \fb's copy engine saturates the link higher than NCCL's
SM kernels. Setup is separate: NCCL needs a $10$--$110$\,s communicator standup;
\fb{} needs none.}
\label{tab:micro}
\begin{tabular}{@{}l c c c c@{}}
\toprule
Collective & 1\,MB & 32\,MB & plateau & plateau BW (fabric vs.\ NCCL) \\
\midrule
1$\to$1 broadcast & 3.33$\times$ & 1.58$\times$ & 1.27$\times$ & 826 vs.\ 651\gbs{} ($+27\%$) \\
1$\to$N broadcast & 4.65$\times$ & 1.72$\times$ & 1.16$\times$ & 740 vs.\ 637\gbs{} ($+16\%$) \\
all-gather        & 2.07$\times$ & 1.60$\times$ & 1.49$\times$ & 740 vs.\ 497\gbs{} ($+49\%$) \\
\bottomrule
\end{tabular}
\end{table}

\noindent The plateaus tell the bandwidth story: \fb{} reaches $826$\gbs{} on a
single link versus NCCL's $651$ ($+27\%$), and on all-gather, where NCCL's
SM-kernel bus bandwidth tops out well below the physical link, $740$ versus $497$
($+49\%$). The small-message regime tells the latency story: at 1\,MB the fabric
copy is $3$--$5\times$ faster because it carries no kernel-launch or protocol
overhead. And neither number includes NCCL's $10$--$110$\,s communicator standup,
which \fb{} eliminates outright (\S\ref{sec:c2}).

\subsection{Single-node FlashLoad}\label{sec:e-load}
\begin{figure}[t]
\centering
\begin{tikzpicture}
\begin{groupplot}[
  group style={group size=2 by 1, horizontal sep=1.7cm},
  width=0.46\textwidth, height=4.6cm, fbbase, ymin=0,
  symbolic x coords={r0,r1,r2,r3}, xtick=data,
  xticklabels={rank0,rank1,rank2,rank3}, enlarge x limits=0.18,
  x tick label style={font=\scriptsize}, title style={yshift=15pt},
  legend style={at={(0.5,1.0)},anchor=south,legend columns=-1,draw=none,font=\scriptsize,
                /tikz/every even column/.append style={column sep=0.35cm}},
]
\nextgroupplot[ybar stacked, bar width=13pt, title={(a) Flash, per-rank load (s)},
  ylabel={time (s)}, ymax=0.62]
\addplot[fill=fbblue,draw=fbblue] coordinates {(r0,0.210)(r1,0.195)(r2,0.236)(r3,0.194)}; \addlegendentry{PIN}
\addplot[fill=fbgreen,draw=fbgreen] coordinates {(r0,0.228)(r1,0.228)(r2,0.222)(r3,0.222)}; \addlegendentry{H2D}
\nextgroupplot[ybar stacked, bar width=13pt, title={(b) Pro, per-rank load (s)},
  ylabel={time (s)}, ymax=2.5]
\addplot[fill=fbblue,draw=fbblue] coordinates {(r0,0.954)(r1,0.968)(r2,0.999)(r3,0.990)};
\addplot[fill=fbgreen,draw=fbgreen] coordinates {(r0,1.183)(r1,1.183)(r2,1.186)(r3,1.187)};
\end{groupplot}
\end{tikzpicture}
\caption{\textbf{Single-node \fl{} critical path} (TP4, daemon-staged,
\code{io\_wait}$=0$). The engine-visible load is PIN$+$H2D only: $\sim$$0.4$\,s per
Flash rank, $\sim$$2.1$\,s per Pro rank; H2D holds $181$--$187$\gbs.}
\label{fig:loadbar}
\end{figure}
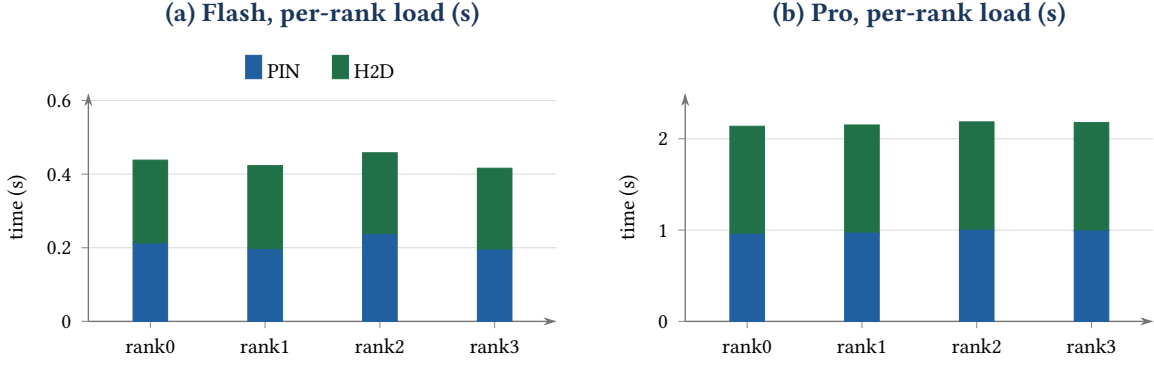

With the cold disk read hidden by the preloader daemon
(\code{io\_wait}$=0$ on all ranks, \S\ref{sec:bulkpipe}), the engine-visible load
collapses to PIN$+$H2D. Figure~\ref{fig:loadbar} shows the per-rank breakdown:
every Flash rank is resident in $\sim$$0.4$\,s and every Pro rank in
$\sim$$2.1$\,s, with H2D running at $181$--$187$\gbs{} (the C2C-bound rate). The PIN
(\code{cudaHostRegister}) cost is not negligible (it is comparable to the H2D), but
both are small and fixed. Table~\ref{tab:load} places these against the baselines:
\fb{} is up to $\sim$$50\times$ faster than the strongest prior loader
(InstantTensor), and up to $77\times$ faster than SafeTensors on Flash from cold
disk, where the baseline pays the full disk-read-on-the-critical-path cost that
\fb{} hides.

\begin{table}[t]
\centering\small
\caption{Single-node load time (s) and \fb{} speedup. Baselines are end-to-end
weight load; \fb{} is the daemon-staged PIN$+$H2D critical path (cold read hidden).}
\label{tab:load}
\begin{tabular}{@{}l l r r r c@{}}
\toprule
Model & Source & SafeTensors & InstantTensor & \fb{} & speedup (vs.\ IT\,--\,ST) \\
\midrule
\multirow{2}{*}{Flash} & cold disk & 30.8 & 20.1 & \textbf{0.4} & $50\times$\,--\,$\mathbf{77\times}$ \\
                       & warm RAM  & 17.7 & 14.8 & \textbf{0.4} & $37\times$\,--\,$44\times$ \\
\multirow{2}{*}{Pro}   & cold disk & 116.8 & 72.9 & \textbf{2.1} & $35\times$\,--\,$56\times$ \\
                       & warm RAM  & 67.4 & 45.8 & \textbf{2.1} & $22\times$\,--\,$32\times$ \\
\bottomrule
\end{tabular}
\end{table}

\subsection{Concurrent multi-node FlashLoad (shard\,+\,all-gather)}\label{sec:e-dp}
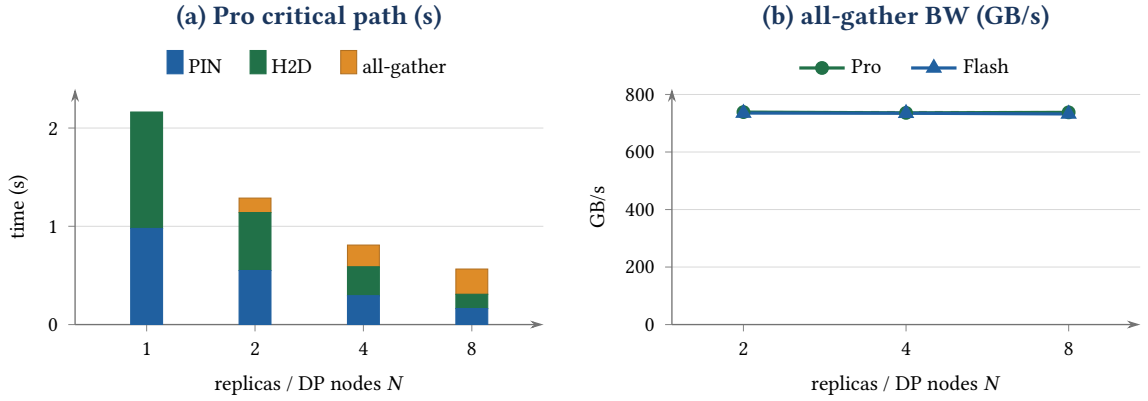
\begin{figure}[t]
\centering
\begin{tikzpicture}
\begin{groupplot}[
  group style={group size=2 by 1, horizontal sep=1.7cm},
  width=0.46\textwidth, height=4.7cm, fbbase, ymin=0,
  symbolic x coords={1,2,4,8}, xtick=data, enlarge x limits=0.22,
  xlabel={replicas / DP nodes $N$}, title style={yshift=12pt},
]
\nextgroupplot[ybar stacked, bar width=12pt, title={(a) Pro critical path (s)},
  ylabel={time (s)}, ymax=2.4,
  legend columns=-1, legend style={at={(0.5,1.0)},anchor=south,draw=none,font=\scriptsize,
    /tikz/every even column/.append style={column sep=0.3cm}},legend cell align=left]
\addplot[fill=fbblue,draw=fbblue] coordinates {(1,0.978)(2,0.55)(4,0.294)(8,0.163)}; \addlegendentry{PIN}
\addplot[fill=fbgreen,draw=fbgreen] coordinates {(1,1.185)(2,0.59)(4,0.294)(8,0.145)}; \addlegendentry{H2D}
\addplot[fill=fbamber,draw=fbamber!80!black] coordinates {(1,0)(2,0.147)(4,0.221)(8,0.257)}; \addlegendentry{all-gather}
\nextgroupplot[title={(b) all-gather BW (GB/s)}, ylabel={GB/s},
  ymin=0, ymax=820, ytick={0,200,400,600,800},
  every axis plot/.append style={very thick,mark size=2pt},
  legend columns=-1, legend style={at={(0.5,1.0)},anchor=south,draw=none,font=\scriptsize,
    /tikz/every even column/.append style={column sep=0.3cm}},legend cell align=left]
\addplot[black!45,dashed,line width=0.6pt,forget plot] coordinates {(2,740)(8,740)};
\addplot[fbgreen,mark=*] coordinates {(2,738.8)(4,736.1)(8,737.9)}; \addlegendentry{Pro}
\addplot[fbblue,mark=triangle*,mark size=2.4pt] coordinates {(2,735.4)(4,734.8)(8,732.3)}; \addlegendentry{Flash}
\end{groupplot}
\end{tikzpicture}
\caption{\textbf{Concurrent multi-node \fl{}} (TP4). \textbf{(a)}~Pro critical path
shrinks $\sim$$2.1$\,s\,$\to$\,$0.57$\,s as PIN/H2D scale $\sim 1/N$ and the ring
all-gather stays modest. \textbf{(b)}~all-gather bandwidth is flat in $N$ and in
model size ($\sim$$732$--$739$\gbs); each GPU reads one predecessor, so there is no
egress contention.}
\label{fig:dp}
\end{figure}

When $N$ replicas come up together, the data-parallel path
(\S\ref{sec:shard}) splits the cold read $N$ ways and reassembles over the fabric.
Figure~\ref{fig:dp}(a) shows the Pro critical path shrinking from $\sim$$2.1$\,s
($N{=}1$, no gather) to $\sim$$0.57$\,s ($N{=}8$): the local PIN and H2D legs each
scale as $\sim 1/N$ (smaller shard per node), while the ring all-gather adds only a
modest, near-constant pass. The off-critical-path cold disk read scales roughly as
$\sim 1/N$ (in fact a little better, since splitting the read also relieves per-node
NVMe contention), from $\sim$$60$\,s for a whole copy to $\sim$$25$\,s ($N{=}2$),
$\sim$$15$\,s ($N{=}4$), and $\sim$$5.7$\,s ($N{=}8$, slowest node), which is the
whole point of the scheme: use the aggregate disk and CPU bandwidth of all nodes.
Figure~\ref{fig:dp}(b) shows the all-gather bandwidth is decoupled from $N$
and from model size: $\sim$$732$--$739$\gbs{} for both Flash and Pro at
$N\in\{2,4,8\}$, because each GPU always reads exactly one predecessor. Adding nodes therefore continues to shrink the critical path, with no bandwidth degradation as $N$ grows.

\subsection{Cross-node FlashClone}\label{sec:e-clone}
\begin{figure}[t]
\centering
\begin{tikzpicture}
\begin{groupplot}[
  group style={group size=2 by 1, horizontal sep=1.7cm},
  width=0.46\textwidth, height=4.8cm, fbbase, title style={yshift=11pt},
  every axis plot/.append style={very thick,mark size=2pt},
]
\nextgroupplot[title={(a) per-clone bandwidth vs.\ $N$}, ylabel={GB/s},
  symbolic x coords={1,2,4,8}, xtick=data, enlarge x limits=0.18,
  xlabel={number of clones $N$}, ymin=0, ymax=900, ytick={0,200,400,600,800},
  legend columns=-1, legend style={at={(0.5,1.0)},anchor=south,draw=none,font=\scriptsize,
    /tikz/every even column/.append style={column sep=0.22cm}},legend cell align=left]
\addplot[fbgreen,mark=*] coordinates {(1,806)(2,722.2)(4,718.5)(8,715.4)}; \addlegendentry{chain, Pro}
\addplot[fbgreen,densely dashed,mark=o] coordinates {(1,804.5)(2,708.9)(4,710.8)(8,704.7)}; \addlegendentry{chain, Flash}
\addplot[fbred,mark=square*,mark size=1.6pt] coordinates {(1,825.1)(2,418)(4,209.2)(8,104.6)}; \addlegendentry{pull (star), Pro}
\nextgroupplot[title={(b) chain-fill BW by position (Pro, $N{=}8$)},
  xlabel={clone rank (chain position)}, ylabel={GB/s},
  xtick={0,1,2,3,4,5,6,7}, ymin=695, ymax=740,
  every axis plot/.append style={very thick,mark size=1.8pt}]
\addplot[fbgreen,mark=*] coordinates {(0,732.5)(1,716.5)(2,714.6)(3,715.5)(4,711.5)(5,713.8)(6,713.6)(7,705.0)};
\end{groupplot}
\end{tikzpicture}
\caption{\textbf{Cross-node \fc.} \textbf{(a)}~Per-clone bandwidth vs.\ clone count:
pull degrades as $\sim$$838/N$ (seed-egress contention); chain stays
flat ($\sim$$715$\gbs{} Pro, $\sim$$705$ Flash). \textbf{(b)}~The only chain cost is
a $\sim$$4\%$ head-to-tail pipeline-fill gradient over the $809$ chunks ($256$\,MiB each) of a Pro
transfer.}
\label{fig:clone}
\end{figure}
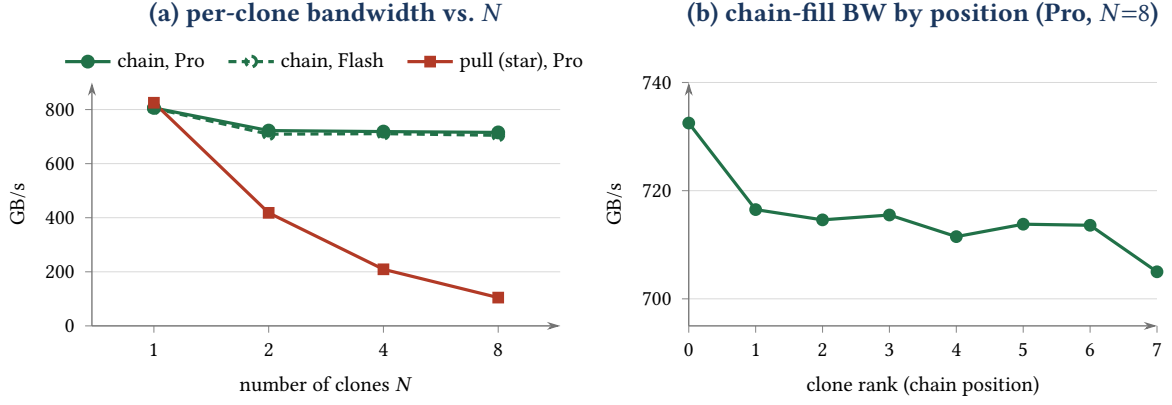

Finally we replicate a resident model seed$\to$clones over the fabric. The fabric
map is $2$--$15$\,ms regardless of payload, replacing NCCL's $10$--$110$\,s.
Figure~\ref{fig:clone}(a) is the central scaling result: under the star
(pull) topology, per-clone bandwidth collapses as $\sim$$838/N$ (Pro:
$825\to418\to209\to105$\gbs{} for $N=1,2,4,8$), because all clones share one seed
GPU's egress; under chain, it stays essentially flat ($806\to715$\gbs{} for
Pro; $805\to705$ for Flash), because each clone reads a distinct predecessor over an
independent link. The two coincide only at $N{=}1$. Chain's advantage over pull
therefore grows with scale: $1.7\times$ at $N{=}2$, $3.4\times$ at $N{=}4$, and
$\mathbf{6.8\times}$ at $N{=}8$. Figure~\ref{fig:clone}(b) shows the only cost chain
pays: a small head-to-tail pipeline-fill gradient ($732\to705$\gbs{} along the
8-deep chain, $\sim$$4\%$), amortized over the $809$ chunks ($256$\,MiB each) of a Pro transfer.

The end-to-end consequence (Table~\ref{tab:clone}): a full Pro replica
($217$\,GB/rank) is resident in $\sim$$0.32$\,s by chain regardless of how many
clones are filled, versus $\sim$$2.1$\,s per clone by pull at $N{=}8$. Set against
the production R-Fork path, which serializes a $10$\,s$+$ NCCL standup and transfer
per clone, bringing up $8$ replicas drops from a serialized
$\gtrsim 8\times10.9\approx87$\,s to a concurrent $\sim$$0.32$\,s, a
$\mathbf{>270\times}$ reduction in time-to-serving for the rack.

\begin{table}[t]
\centering\small
\caption{\fc{} end-to-end per-clone cost ($N{=}8$): a one-time map plus the
transfer. Pull's transfer balloons under shared egress; chain holds near the
single-link rate. (Pro $217$\,GB/rank, Flash $41.4$\,GB/rank.)}
\label{tab:clone}
\begin{tabular}{@{}l l r r r@{}}
\toprule
Model & Topology & map (ms) & transfer (ms) & per-clone BW \\
\midrule
\multirow{2}{*}{Pro}   & pull  & 7--13 & 2075 & 104.6\gbs \\
                       & \textbf{chain} & 7--13 & \textbf{303} & \textbf{715.4}\gbs \\
\multirow{2}{*}{Flash} & pull  & 2--6 & 395 & 104.9\gbs \\
                       & \textbf{chain} & 3--7 & \textbf{59} & \textbf{704.7}\gbs \\
\bottomrule
\end{tabular}
\end{table}

\section{Discussion and Ablations}\label{sec:disc}

\subsection{Chunk size: the chain's one knob}
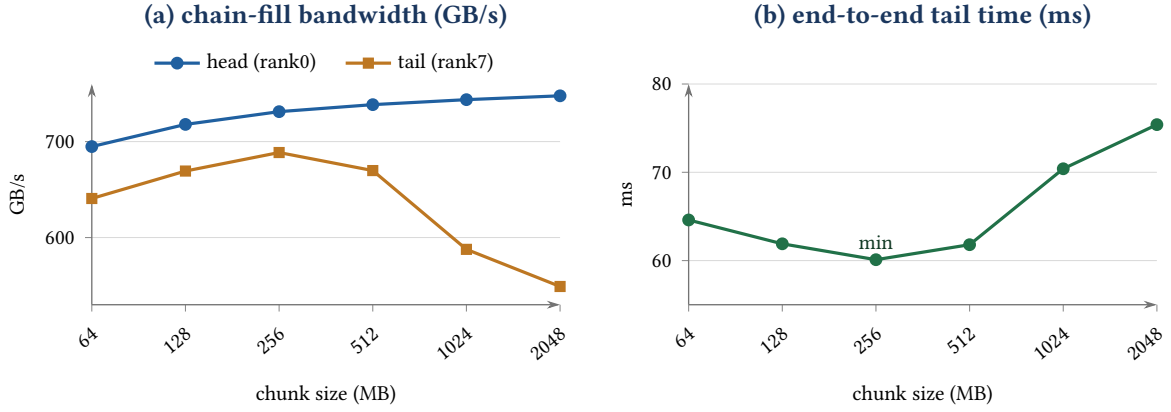
\begin{figure}[t]
\centering
\begin{tikzpicture}
\begin{groupplot}[
  group style={group size=2 by 1, horizontal sep=1.7cm},
  width=0.46\textwidth, height=4.5cm, fbbase,
  xmode=log, log basis x=2, xtick={64,128,256,512,1024,2048},
  xticklabels={64,128,256,512,1024,2048},
  x tick label style={font=\scriptsize,rotate=40,anchor=north east},
  xlabel={chunk size (MB)}, title style={yshift=11pt},
  every axis plot/.append style={very thick,mark size=1.8pt},
]
\nextgroupplot[title={(a) chain-fill bandwidth (GB/s)}, ylabel={GB/s},
  ymin=530, ymax=760,
  legend columns=-1, legend style={at={(0.5,1.0)},anchor=south,draw=none,font=\scriptsize,
    /tikz/every even column/.append style={column sep=0.3cm}},legend cell align=left]
\addplot[fbblue,mark=*] coordinates {(64,694.8)(128,717.9)(256,731.2)(512,738.5)(1024,743.7)(2048,747.7)}; \addlegendentry{head (rank0)}
\addplot[fbamber!85!black,mark=square*,mark size=1.5pt] coordinates {(64,640.7)(128,669.3)(256,688.6)(512,669.8)(1024,587.7)(2048,549.0)}; \addlegendentry{tail (rank7)}
\nextgroupplot[title={(b) end-to-end tail time (ms)}, ylabel={ms},
  ymin=55, ymax=80]
\addplot[fbgreen,mark=*] coordinates {(64,64.6)(128,61.9)(256,60.1)(512,61.8)(1024,70.4)(2048,75.4)};
\node[font=\scriptsize,color=fbgreen!50!black,anchor=south] at (axis cs:256,60.1) {min};
\end{groupplot}
\end{tikzpicture}
\caption{\textbf{Chunk-size sweep (Flash chain).} \textbf{(a)}~the head bandwidth
rises with chunk size, but the tail peaks near $256$\,MB. \textbf{(b)}~the
end-to-end tail time is minimized at \textbf{256\,MB}: smaller chunks add per-chunk
overhead, larger chunks deepen the fill tail.}
\label{fig:chunk}
\end{figure}

The pipelined chain (\S\ref{sec:topo}) has a single tunable, the chunk size, that
trades two effects. Larger chunks let each copy-engine DMA run closer to peak, so
the head of the chain speeds up monotonically; but they also deepen the
fixed pipeline-fill latency, which lengthens the tail. Figure~\ref{fig:chunk}
sweeps a Flash chain transfer: the head bandwidth rises steadily from $695$ to
$748$\gbs{} as the chunk grows $64\to2048$\,MB, but the end-to-end tail (the figure
that actually decides ``how long until the last clone is ready'') is minimized at
\textbf{256\,MB} ($60.1$\,ms). Below it, per-chunk synchronization overhead grows;
above it, the fill tail dominates. We therefore default to $256$\,MB, which also
matches the large-transfer optimum seen independently for Pro.

\subsection{All-gather scheduling: predecessor-only ring vs.\ staggered pulls}
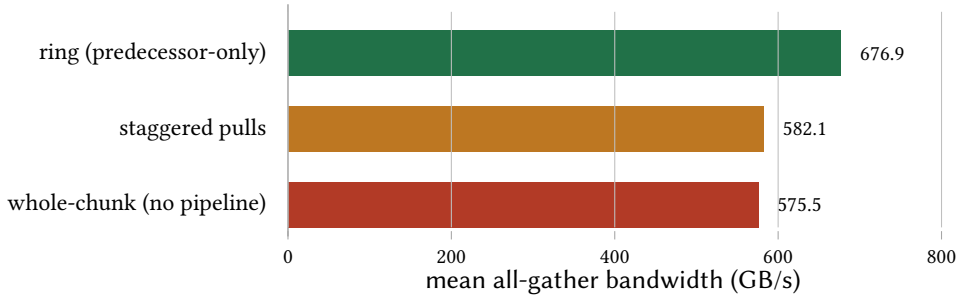
\begin{figure}[t]
\centering
\begin{tikzpicture}[font=\small]
\def\sx{0.0108}
\foreach \y/\v/\c/\lab in {%
  2/676.9/fbgreen/ring (predecessor-only),%
  1/582.1/fbamber!85!black/staggered pulls,%
  0/575.5/fbred/whole-chunk (no pipeline)}{
  \fill[\c] (0,\y) rectangle (\v*\sx,\y+0.6);
  \node[anchor=east,font=\footnotesize,color=black] at (-0.15,\y+0.3) {\lab};
  \node[anchor=west,font=\scriptsize,color=black] at (\v*\sx+0.12,\y+0.3) {\v};
}
\draw[black!40,line width=0.5pt] (0,-0.12)--(0,2.95);
\foreach \t in {0,200,400,600,800}{
  \draw[black!25,line width=0.4pt] (\t*\sx,0)--(\t*\sx,2.85);
  \draw[black!35] (\t*\sx,-0.05)--(\t*\sx,-0.16) node[below,font=\scriptsize,color=black]{\t};}
\node[font=\small\sffamily,color=black] at (4.3,-0.72) {mean all-gather bandwidth (GB/s)};
\end{tikzpicture}
\caption{\textbf{All-gather scheduling ablation} (Pro, $N{=}8$). Predecessor-only
ring beats staggered pulls (more uniform: $\pm$$31$ vs.\ $\pm$$118$\gbs{} spread;
import $6$--$9$ vs.\ $30$--$55$\,ms), and sub-chunk pipelining beats a whole-chunk
store-and-forward ring.}
\label{fig:ring}
\end{figure}

The data-parallel all-gather (\S\ref{sec:shard}) can be scheduled two ways. A
staggered schedule has each node directly pull the other $N{-}1$ owned ranges
in a rotating permutation; a ring has each node read only from its predecessor and
forward onward, overlapped via flag-pipelined sub-chunks. The ring
wins on three counts (Figure~\ref{fig:ring}): higher mean bandwidth, far more
uniform per-node bandwidth (because no node is ever read by more than one peer), and
an order-of-magnitude cheaper handle import (each node imports one predecessor, not
$N{-}1$ peers: $6$--$9$\,ms vs.\ $30$--$55$\,ms). The figure also shows why
sub-chunk pipelining matters: a whole-chunk ring (one $\sim$$27$\,GB copy per hop,
no pipelining) is store-and-forward and leaves links idle, costing both mean
bandwidth and uniformity.\footnote{The absolute bandwidths in
Figure~\ref{fig:ring} are from the scheduling ablation runs (256\,MB sub-chunks);
the production byte-equal ring of \S\ref{sec:e-dp} reaches $\sim$$737$\gbs{} with a
whole-shard chunk, the configuration we ship. The relative ordering
(ring\,$>$\,staggered\,$>$\,whole-chunk) is the point here.}

\subsection{Beyond NVL72: NVLink islands and RDMA}\label{sec:beyond}
\fb's only hard requirement is a primitive for one node to grant another direct
read access to a memory region, plus a small out-of-band channel for the access
token. On NVL72 we realize this with the CUDA VMM fabric handle
(\code{CU\_MEM\_HANDLE\_TYPE\_FABRIC}) and an IMEX channel, the internode
memory-sharing service of multi-node NVLink platforms (\S\ref{sec:remotemap}). It
helps to separate which parts of the design are platform-independent from the two
that our implementation ties to that fabric. The contiguous arena is
platform-independent: laying weights out as one large 1-D image addressed by
offsets pays off on \emph{any} interconnect, because a single bulk transfer
saturates a link that tens of thousands of per-tensor objects cannot (C1),
whether the mover is a copy engine, a GPUDirect P2P read, or an RDMA verb. The
fabric-specific pieces are exactly the two that cross nodes: the fabric-handle
export/import, which goes through IMEX and so does not span an RDMA boundary, and
the \code{cudaMemcpy} device-to-device mover, which reaches a remote GPU only
because IMEX has already mapped that GPU's memory into the local address space.
Neither survives a move to a platform whose nodes are joined by RDMA rather than
NVLink.

Table~\ref{tab:h100} makes this concrete for an H100-class deployment whose nodes
are internally NVSwitch-connected and externally joined by RDMA.
\emph{Within} a node the design runs as built: the arena, the bulk H2D load,
zero-copy serving, and the \code{cudaMemcpy} chain/ring movers all operate over
intra-node NVLink P2P (the one adjustment is the cross-process share handle, which
on a node without an IMEX/MNNVL fabric becomes a POSIX file-descriptor IPC handle
rather than a fabric handle). \emph{Across} nodes the fabric map and the
\code{cudaMemcpy} mover do not apply, so cross-node \fc{} and the multi-node \fl{}
all-gather, as implemented, do not run over RDMA. Closing that gap is a backend,
not a redesign: register each arena once, exchange the memory key out of band, and
issue one-sided RDMA reads in place of \code{cudaMemcpy}, so the contiguous layout,
NCCL-free mapping, and chunk-pipelined chain carry over with RDMA bandwidth in
place of NVLink's. An RDMA backend does, however, need the connection machinery
NVLink let us skip: a warm, reusable communicator or queue-pair pool so the
one-time standup is \emph{cached} across scale-out events instead of re-paid each
time (NCCL's lazy connection establishment and \code{ncclCommSplit} reuse, plus its
scalable bootstrap, are the relevant primitives~\cite{ncclinit}), and a
GPUDirect/RDMA pipelined broadcast to play the role our chain does. That broadcast
is well precedented: pipelined and binomial broadcasts over GPUDirect RDMA are
standard in MPI~\cite{gdrbcast}, and recent fast-scaling systems for LLM inference
move weights this way, e.g.\ $\lambda$Scale's block-pipelined one-sided-read
broadcast~\cite{lambdascale} and TransferEngine's pipelined one-sided
writes~\cite{transferengine}. A cross-platform study with such an RDMA backend on
B300/H100 islands is left for future work; we expect the relative advantages to
hold and the absolute numbers to track each platform's link bandwidth.

\begin{table}[t]
\centering\small
\caption{\fb{} on an H100-class platform (intra-node NVSwitch, inter-node RDMA),
component by component. Marks: \textcolor{fbgreen}{\ding{51}}~works as implemented;
\textcolor{fbamber!85!black}{partial}~needs a small change (e.g.\ a POSIX-FD share
handle); \textcolor{fbred}{\ding{55}}~not supported by the current implementation.
The contiguous layout is platform-independent; only the cross-node movers are
NVLink/IMEX-specific.}
\label{tab:h100}
\begin{tabular}{@{}l c c@{}}
\toprule
Component & Intra-node (NVSwitch) & Inter-node (RDMA) \\
\midrule
\fa{} contiguous layout $+$ zero-copy views & \textcolor{fbgreen}{\ding{51}} & \textcolor{fbgreen}{\ding{51}} \\
\fl{} bulk H2D load (CPU$\to$GPU)           & \textcolor{fbgreen}{\ding{51}} & \textcolor{fbgreen}{\ding{51}} \\
Fabric handle export/import (FABRIC\,+\,IMEX) & \textcolor{fbamber!85!black}{partial} & \textcolor{fbred}{\ding{55}} \\
\code{cudaMemcpy} remote map \& D2D clone   & \textcolor{fbgreen}{\ding{51}} & \textcolor{fbred}{\ding{55}} \\
Chain/ring pipelined broadcast (as built)   & \textcolor{fbgreen}{\ding{51}} & \textcolor{fbgreen}{\ding{51}} \\
Shard\,+\,all-gather DP load (as built)     & \textcolor{fbgreen}{\ding{51}} & \textcolor{fbgreen}{\ding{51}} \\
Chain/all-gather over an RDMA backend       & ---                            & \textcolor{fbgreen}{\ding{51}} \\
\bottomrule
\end{tabular}
\end{table}

\subsection{Limitations}
A few caveats bound the results. The seed must keep its arena resident for the
service lifetime; if it exits, handles go stale and clones re-fetch. All
participants must share an IMEX channel as the same user (a per-user security
model), with a graceful fall-back to NCCL when IMEX is unavailable. The
contiguous-arena scheme assumes the expert/non-expert separator stays aligned;
both evaluation models satisfy this, but a future model that violates it would need
a padded separator (and a second H2D for the padded region). Finally, the results
are scoped to what we have built out, not what the design admits, along two axes. In
software, the \fl{} fast path targets \code{DeepseekV4ForCausalLM} under TP and EP
(with DP and DP-attention layouts captured by the pre-pack), so other MoE families,
pipeline parallelism, and the cross-node RDMA backend are the extensions identified
in \S\ref{sec:impl} and \S\ref{sec:beyond}. In hardware, every number reported here
is measured on the GB300 NVL72; the portability to the wider rack-scale class (Vera
Rubin POD, AMD Helios, Huawei CloudMatrix384, TPU pods) rests on the
platform-independent contiguous arena and is argued structurally (\S\ref{sec:beyond}),
not yet measured.

\section{Conclusion}\label{sec:concl}
In elastic deployment of large MoE models, weight loading is a cost worth
optimizing, and we showed that on a rack-scale GB300 NVL72 that cost is structural,
not a bandwidth shortfall: fragmented per-tensor memory starves the interconnect,
NCCL communicator setup dwarfs the transfer it gates, and the production
replication path is serial. \fb{} addresses all three with one coherent co-design:
make the weights one contiguous, inter-node--exportable arena and serve them in
place, then load from CPU as a single bulk zero-copy transfer with the cold read
hidden (\fl), and replicate GPU$\to$GPU by mapping remote memory in $\sim$$10$\,ms,
with no NCCL, and broadcasting along a chain that scales flat in the number of clones
(\fc). The result is up to $\sim$$50\times$ faster single-node weight loading,
$\geq$$700$\gbs{} per-clone replication, a Pro replica resident in $\sim$$0.32$\,s,
and $>270\times$ faster concurrent multi-node weight loading than the state of the
art. As the rack-scale system becomes the industry's unit of large-model deployment,
we expect these principles, layout first, map don't communicate, and no
single-point bottleneck, to carry across the emerging platforms (\S\ref{sec:beyond})
and to remain a foundation for fast model loading on any memory-mappable accelerator
fabric.


\end{document}